\documentclass[12pt,preprint]{aastex}

\slugcomment{Submitted to the Astrophysical Journal}

\shorttitle{Fragmentation of a Molecular Cloud Core}
\shortauthors{Matsumoto and Hanawa}

\begin{document}

\title{Fragmentation of a Molecular Cloud Core versus Fragmentation of
the Massive Protoplanetary Disk in the Main Accretion Phase}

\author{Tomoaki Matsumoto\altaffilmark{1,2}}
\email{matsu@i.hosei.ac.jp}

\and

\author{Tomoyuki Hanawa\altaffilmark{3}}
\email{hanawa@cfs.chiba-u.ac.jp}

\altaffiltext{1}{Department of Humanity and Environment, Hosei University,
 Fujimi, Chiyoda-ku, Tokyo 102-8160, Japan}
\altaffiltext{2}{Theoretical Astrophysics, National Astronomical
 Observatory, Mitaka, Tokyo 181-8588, Japan}
\altaffiltext{3}{Center for Frontier Science, Chiba University, 
Yayoi-cho, Inage-ku, Chiba, 263-8522, Japan}

\begin{abstract}

The fragmentation of molecular cloud cores a factor of 1.1 denser than
the critical Bonnor-Ebert sphere is examined though three-dimensional
numerical simulations. A nested grid is employed to
resolve fine structure down to 1~AU while following the entire structure of
the molecular cloud core of radius 0.14~pc.
A barotropic equation of state is assumed to take account of the change in
temperature during collapse, allowing simulation of the formation of
the first core.  A total of 225 models are shown 
to survey the effects of initial rotation speed, rotation law, and
amplitude of bar mode perturbation. The simulations show that the
cloud fragments whenever the cloud rotates sufficiently slowly to allow
collapse but fast enough to form a disk before first-core
formation.  The latter condition is equivalent to $\Omega_0 t_{\rm ff}
\gtrsim 0.05$, where $\Omega_0$ and $t_{\rm ff}$ denote the initial
central angular velocity and the freefall time measured from the central
density, respectively.  Fragmentation is classified into six types:
{\it disk-bar}, {\it ring-bar}, {\it satellite}, {\it bar}, {\it
ring}, and {\it dumbbell} types according to the morphology of
collapse and fragmentation.  
When the outward decrease in initial angular velocity is more steep, the cloud
deforms from spherical at an early stage.  The cloud deforms into a ring only
when the bar mode ($ m = 2 $) perturbation is very minor.
The ring fragments into two or three fragments via {\it ring-bar} type fragmentation
and into at least three fragments via {\it ring} type fragmentation.  When the bar
mode is significant, the cloud fragments into two
fragments via either {\it bar} or {\it dumbbell} type fragmentation.
These fragments eventually merge due to their low angular
momenta, after which several new fragments form around the
merged fragment via {\it satellite} type fragmentation. 
This satellite type fragmentation may be responsible for observed wide
range of binary separation.

\end{abstract}


\section{Introduction}

It is widely accepted that binary and multiple stars form as a result
of fragmentation in collapsing molecular cloud cores
\citep[e.g.,][]{Bodenheimer2000}.  In the last two decades, 
fragmentation of the molecular cloud core has been investigated through
numerical simulations by many authors.  Criteria for the fragmentation of
isothermally collapsing clouds has been investigated by
\citet*{Miyama84,Boss93,Boss95,Tsuribe99}.  Their criteria have
converged as far as the isothermal phase is concerned; 
a cloud with $\alpha \lesssim 0.2 - 0.5$ collapses into
fragments depending little on $\beta$, even when the initial density
and velocity distributions differ. Here, $\alpha$ and $\beta$ denote
the ratio of thermal energy to gravitational energy and that of
rotation energy to gravitational energy, respectively.
\citet{Tsuribe99} pointed out that the criteria for fragmentation
corresponds to the formation of a flat disk with flatness greater than $4
\pi$.  On the other hand, a cloud with $\alpha \gtrsim 0.2 - 0.5$
collapses self-similarly and shows no sign of
fragmentation. \citet{HM99} and \citet{MH99} investigated deformation
of the self-similarly collapsing cloud in search of the possibility
that deformation of the central cloud to a bar might trigger
fragmentation. The growth of the bar mode is slow compared to the
timescale of the collapse, i.e., $\Delta \propto \rho_{\rm
max}^{0.177}$, where $\Delta$ and $\rho_{\rm max}$ denote the
amplitude of the bar mode and the maximum density of the cloud,
respectively.  The bar may indeed fragment, but only at a later stage.

As the collapse proceeds, the cloud core becomes optically 
thick and the efficiency of radiative cooling 
decreases.   The temperature starts increasing 
when the central density exceeds 
the critical density of $\sim 10^{-13}\,{\rm g}\,{\rm cm}^{-3}$.
This increase in temperature results in the formation of a quasi-static
core, i.e., the first core of \citet{larson69}.
The first core grows by accreting gas, and this accretion phase
persists for long enough for the core to fragment.
Thus, the dynamics of the cloud changes qualitatively at
the critical density.  Stability against fragmentation is
also likely to change at this critical density.
In fact, the first core has been shown by recent simulations to be very
unstable taking account of the change in temperature
\citep*{Burkert97,Nelson98,Sigalotti98,Klapp98,Boss00}.
Their simulations, however, assume rather small $ \alpha $.
When $ \alpha $ is small, the cloud is Jeans unstable
and fragments easily \citep{Tsuribe99}.
The fragmentation of the first core may be due to the
small $ \alpha $ assumed.
It is still unanswered whether the first core fragments
when the initial cloud has a moderately large
$ \alpha $.  
Thus, we investigate the fragmentation of the cloud with focus on the cloud with
large $\alpha$ of 0.765.  The initial cloud is
only 1.1 times more massive than the critical Bonnor-Ebert sphere
\citep{Ebert1955,Bonnor1956}, which is an equilibrium state of the
isothermal cloud. 

The Bonnor-Ebert sphere provides a good fit to the density distribution
of a dark globule.  According to recent near-infrared observations
\citep*{Alves01}, the model with $\xi = 6.9 \pm 0.2$ gives the best fit
for B68, where $\xi$ denotes the non-dimensional radius.  Similarly,
those with $\xi = 12.5 \pm 2.6$ and $7.0 \pm 0.3$ give a good fit for
B335 \citep{Harvey01} and the Coalsak \citep{Racca02}, respectively.
When $\xi > 6.45$, the Bonnor-Ebert sphere is unstable against
collapse. Thus, our initial model can be applied to these globules.

The model can also be applied to cloud cores embedded with 
molecular clouds.  The masses of such cores evaluated from C$^{18}$O luminosity
are similar to the virial masses \citep[e.g.,][]{onishi96}. This
implies that the cores are gravitationally bound and nearly in equilibrium,
and accordingly, the parameter $\alpha$ is only slightly less than unity.


Recent numerical studies have lacked any survey of
model parameters. In this study, 225 models
with different rotation speed, rotation law, and amplitude of bar mode
perturbation are considered. The simulations show many types of fragmentation,
some of which are new.
The main features of each type and 
the territory of each type in the parameter space are discussed.

In this paper, the collapse and fragmentation of
molecular cloud cores is investigated using a nested grid. 
The nested grid has high spatial resolution near the
center of the computation domain and allows 
fragmentation to be followed without violating 
the Jeans condition \citep{Truelove97}. 
In \S \ref{sec:models}, the models of cloud cores are introduced.
In \S \ref{sec:methods}, the methods of numerical simulations are presented. 
In \S \ref{results}, the results are shown and the fragmentation is classified.
In \S \ref{sec:discussion}, the origins of different types of
fragmentation are discussed, the simulations are compared with 
earlier numerical works, the implications on binary formation are related. 
The paper is concluded in \S \ref{sec:summary}.

\section{Models}
\label{sec:models}
As a model for molecular cloud cores, 
we consider Bonnor-Ebert spheres, which 
belong to a sequence of equilibrium-state
isothermal spherical clouds confined by external pressure
\citep{Ebert1955,Bonnor1956}. 
Given the external pressure ($ P _{\rm ex} $) and 
the sound speed ($ c _s $), 
the Bonnor-Ebert sphere is stable against collapse 
only when the central density is lower than the critical value,
$  14.0\, P _{\rm ex} / c _s^2 $.
The critical Bonnor-Ebert sphere is used as a template for 
the model clouds examined in this study.   In the models, the initial density 
distribution is given by 
\begin{equation}
\rho(r) = \rho _c \varrho _{\rm BE} (r/a) 
\end{equation}
and
\begin{equation}
a = c _s \left(\frac{f}{4 \pi G \rho _c}\right)^{1/2} \; ,
\end{equation}
where $r$, $f$, $ \rho _c $ and $G$ denote
the radius, density enhancement factor, the initial central density, 
and the gravitational constant,
respectively.  The function $\varrho_{\rm BE}$ denotes
the density distribution of the critical
Bonnor-Ebert sphere, and can be approximated as
\begin{equation}
\varrho _{\rm BE} (\xi) = 1 - \frac{\xi ^2}{6} + \frac{\xi^4}{45} +
{\cal O}(\xi^6)\;.
\end{equation}
The critical Bonnor-Ebert sphere has radius of
$ \xi = 6.45 $.  A density enhancement factor of 
$f = 1.1$ is assumed in typical models because observed molecular clouds
are nearly in the virial equilibrium.
This slight density enhancement collapses rotating clouds when
the initial cloud rotates slowly.

The initial central density is set at 
$\rho_c = 1 \times10^{-19}\,{\rm g}\,{\rm cm}^{-3}$, 
which corresponds to a number density of
$n_c = 2.6 \times10^4 \,{\rm cm}^{-3}$ for the assumed
mean molecular weight of 2.3.
An initial temperature of $T=10\,{\rm K}$ is assumed, and
hence $ c_s = 0.19 \,{\rm km}\,{s}^{-1}$.
The radius and mass of the cloud are thus 
$R_c=0.144\;{\rm pc}$ and $M=3.24\;M_\odot$ for $ f = 1.1 $.

The initial velocity includes only the $ \varphi $-component, and
the angular velocity depends on $ R $ and $ \varphi $ in
cylindrical  coordinates $ (R, \varphi, z) $. 
The angular velocity $ \Omega $ is expressed as 
\begin{equation}
\Omega(R,\varphi) = \left[\Omega_0 + \Omega_2 \cos \left(2 \varphi\right) 
+ \Omega_3 \left( \frac{R}{a} \right) \cos\left(3 \varphi\right)\right]
\left[ 1 +  2 C \left(\frac{R}{a}\right)^2 \right]^{-1/2} \; ,
\label{eq:velocity}
\end{equation}
where $ \Omega _0 $ denotes the amplitude of the global rotation, and 
$ \Omega _2 $ and $ \Omega _3 $ denote the amplitudes of the
velocity perturbation of $m=2$ and 3.
The parameter $ C $ specifies the dependence of $\Omega$ on $R$.
When $ C = 0 $, the angular velocity is
independent of $ R $ and rotation is ``rigid''.
When $ C $ is larger, the angular
velocity decreases more rapidly with increasing in $ R $.
As shown later, fragmentation of the cloud core depends strongly on 
the parameter $ C $.  
A small amplitude for the perturbation of $m=3$ is set,
such as $\Omega_3 t_{\rm ff} = 1 \times 10^{-3}$, in all the models, where 
$t_{\rm ff}$ denotes the initial freefall timescale at the center and is
defined as $(3 \pi / 32 G \rho_c)^{1/2}$.  This $ m $ = 3 mode
breaks the point symmetry with respect to $ R $ = 0, and accordingly
the fragments are slightly asymmetric in this simulation.
The model parameters $\Omega_0$, $\Omega_2$, and $C$ are varied to
investigate the effects of rotation speed, amplitude of the bar mode,
and the rotation law on fragmentation of the cloud
cores. 

\begin{figure}
\epsscale{0.9}
\plotone{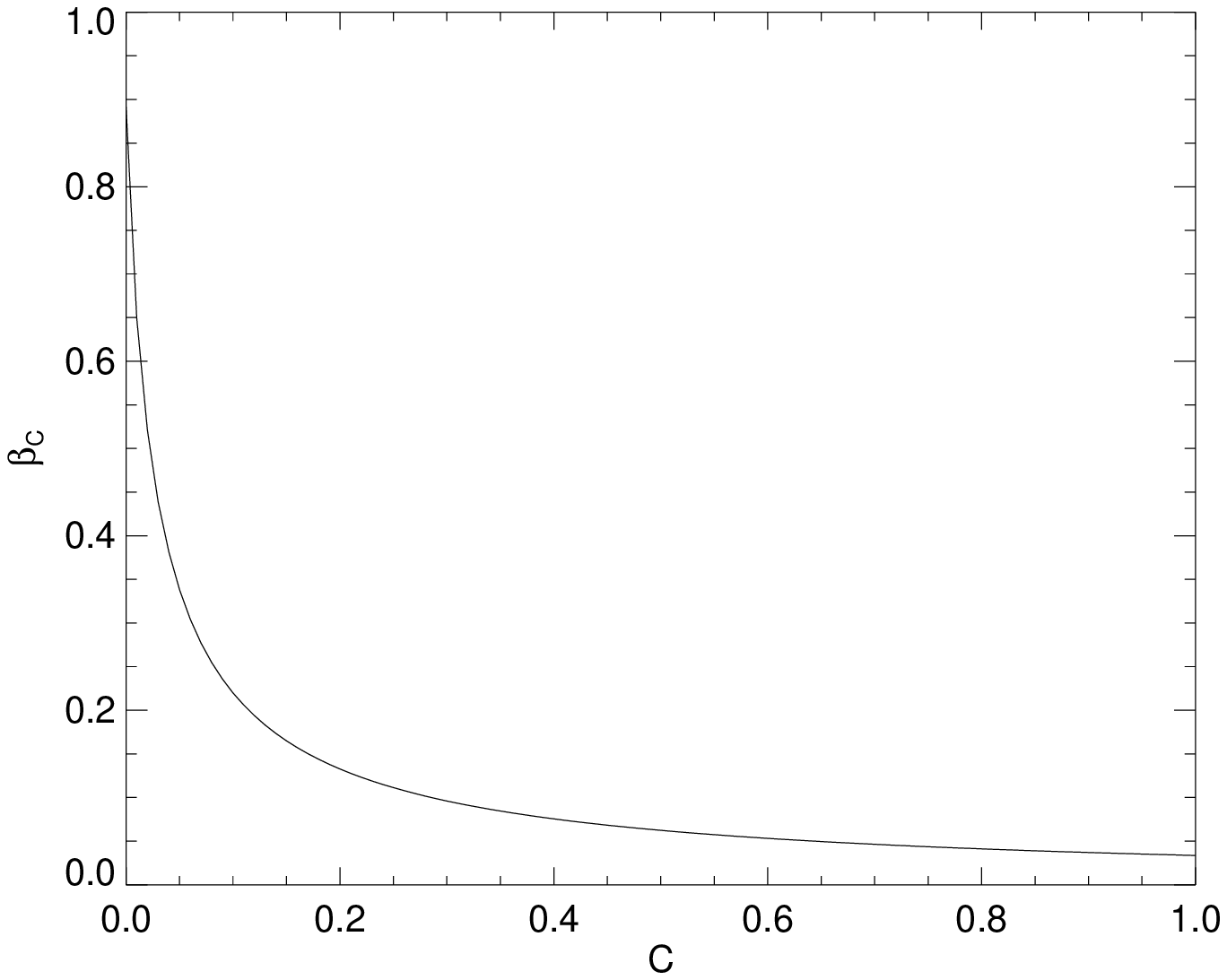}
\figcaption[c-beta-plot.eps]{
Coefficient $\beta_C $ as a function of $C$.
\label{c-beta-plot.eps}
}
\end{figure}

To compare our initial models with those of earlier
simulations, the ratios
$\alpha = E_{\rm th}/|E_{\rm grav}|$
and 
$\beta = E_{\rm rot}/|E_{\rm grav}|$ are computed, 
where $E_{\rm th}$, $E_{\rm rot}$,  and $E_{\rm grav}$ are
thermal energy, rotation energy,  and gravitational energy,
respectively  \citep[e.g.,][]{Tohline81,Miyama84}.
In our model, the initial cloud has $\alpha = 0.765 \, (1.1/f) $. 
The parameter $\beta$ is independent of $f$, and
when $\Omega_3 = 0$, 
$\beta$ is described by
\begin{equation}
\beta = \beta_C t_{\rm ff}^2 \left(
\Omega_0  ^2 +\frac{1}{2} \Omega_2^2 \right)\;,
\label{eq:beta}
\end{equation}
where the coefficient $\beta_C$ is a
function of $C$ as shown in Figure \ref{c-beta-plot.eps}. 
When a cloud rotates rigidly ($C=0$), it has $\beta_C=0.892$. 
When $C\sim1$, $\beta_C$ decreases approximately
in proportion to $C^{-1/2}$.  The decrease in $\beta_C$ is due to slow
rotation in the outer part of the cloud. 

When $ \alpha $ and $ \beta $ are small,
the cloud is unstable against fragmentation.  When the cloud
has uniform density and rotates rigidly, the criterion for 
fragmentation is $ \alpha \lesssim 0.5 $  
\citep[e.g.,][]{Tsuribe99}.  When the cloud is centrally
peaked and the axis ratio is 1.5, the criterion is $ \alpha \lesssim 0.45 $ 
for low $ \beta $ \citep{Boss93}.
When the cloud is more oblate, i.e., the axis ratio is 2.0,
the criterion is $ \alpha \lesssim 0.33 $ for low $ \beta $.
This suggests that a cloud of $ \alpha \gtrsim 0.5 $ 
is stable against fragmentation.  In this study, the possibility of
fragmentation of a cloud with $ \alpha > 0.7 $ is examined.

The dynamical evolution of a cloud is followed taking account of
the self-gravity and gas pressure.  The magnetic field is neglected for
simplicity. 
The gas temperature is
assumed to be 10~K below the critical density
$\rho_{\rm cr} = 2 \times 10^{-13}\,{\rm g}\,{\rm cm}^{-3}$
($ n _{\rm cr} = 5.24\times10^{10}\,{\rm
cm}^{-3} $), and to increase adiabatically in proportion to 
$ \rho ^{2/5} $ above it. 
In other words, a barotropic equation of state is assumed, as expressed by
\begin{equation}
P = \left\{ 
\begin{array}{ll}
c_s^2 \rho &                       {\rm for}\; \rho < \rho_{\rm cr}\\
c_s^2 \rho_{\rm cr} (\rho/\rho_{\rm cr})^{7/5} & {\rm for}\; \rho \geq \rho_{\rm cr}
\end{array}
\right.\;.
\end{equation}
This change in temperature reproduces the
formation of the adiabatic core, which corresponds to
the first core of \citet{larson69}.
The value of the critical density $\rho_{\rm cr}$ is taken from 
the numerical results of \citet{Masunaga98}, who studied the spherical
collapse of molecular cloud cores with radiative hydrodynamics.

\section{Numerical Methods}
\label{sec:methods}
In the simulations, the hydrodynamical equation and Poisson equation
are solved by a finite difference method with 
second-order accuracy in time and space.  A nested grid is employed to
solve the central region with higher spatial resolution.
The hydrodynamic code for the nested grid was developed by
extending the simulation code of \citet{MH99}.
The nested grid consists of concentric hierarchical 
rectangular grids \citep*{York1993}, and the cell width of each grid decreases 
successively by a factor of two.  In the following, 
the coarsest grid is labeled level $ l $ = 1.   
The $ l $-th level grid has $ 2 ^{l -1} $ times higher 
spatial resolution than the coarsest grid.
All the fluxes are conserved at the interface between the coarse
and fine grids as in the standard
adaptive mesh refinement \citep[AMR;][]{chiang1992}.  Thus, the total
mass is conserved in our computations.
The numerical fluxes are obtained by the method of 
\citet{Roe1981} with modification to
solve the isothermal and polytrope gas. 
A MUSCL approach and predictor-corrector method are adopted for time integration
\citep[e.g.,][]{Hirsch1990}. 
The Poisson equation is solved by a multigrid iteration on a nested gird
\citep{MH02}.  This code solves self-gravity consistently over all
grids with different levels such that ``the gravitational field line''
is continuous at interfaces between coarse and fine grids. 
This consistency ensures that the obtained gravitational 
potential is accurate at least up to the quadrapole moment of 
a binary.  Thus, the gravitational torque induced by a binary 
is accurately taken into account in our simulation.

Mirror symmetry with respect to the $ z = 0 $ 
plane is employed to reduce computation cost. A fixed boundary
condition is set for the surface of $r=R_c$, representing
a constant external pressure that confines the cloud during evolution.  
Gas is considered only in $r \leq R_c$ when solving the Poisson equation.

In this paper, each grid has 
$ 256\times 256 \times 32 $ cubic cells in high-resolution models,
and $ 128\times 128 \times 16 $ cubic cells in low-resolution models in $(x, y, z)$.
The model parameters of the high-resolution models are shown in
Table \ref{table:model}. 
The other models shown in this paper are the low-resolution models.
The nested grid consists of grids of 5 levels  at the initial stage.
A new finer grid is introduced to maintain the Jeans condition of 
$ \lambda _{\rm J}/4 > h $ with ample
margin \citep{Truelove97}, where $ \lambda _{\rm J} $
and $ h $ are the Jeans length and cell width, respectively.  
Whenever an eighth of 
the minimum Jeans length ($ \lambda _{\rm J, min} $)
became smaller than the cell width in the finest grid, a new finer grid was
added to the nested grid. 
This means that a finer grid was added with ample margin of
factor 2.
Typical models have 14 grid levels at the last stage.
The Jeans condition was only violated in 
these simulations when a high density fragment escaped from the region covered
by the finest grid, and the computation was terminated in the stage
that this occurred. 
In the model shown in \S\ref{sec:disk},
evolution was successfully computed up to the stage in which the mass of an adiabatic core
(total mass in the region of $\rho \geq \rho_{\rm cr}$) reached 
0.07 $ M_\odot $.

\section{Results}
\label{results}
\subsection{Rigidly Rotating Cloud}
In this subsection,
a total of 27 models of a rigidly rotating cloud ($C=0$) in the region 
$0.03\leq \Omega_0 t_{\rm ff} \leq 0.3$
and 
$0 \leq \Omega_2 t_{\rm ff} \leq 0.3$ are presented
to study the dependence on $\Omega_0$ and $\Omega_2$. 

\begin{figure}
\epsscale{0.9} 
\plotone{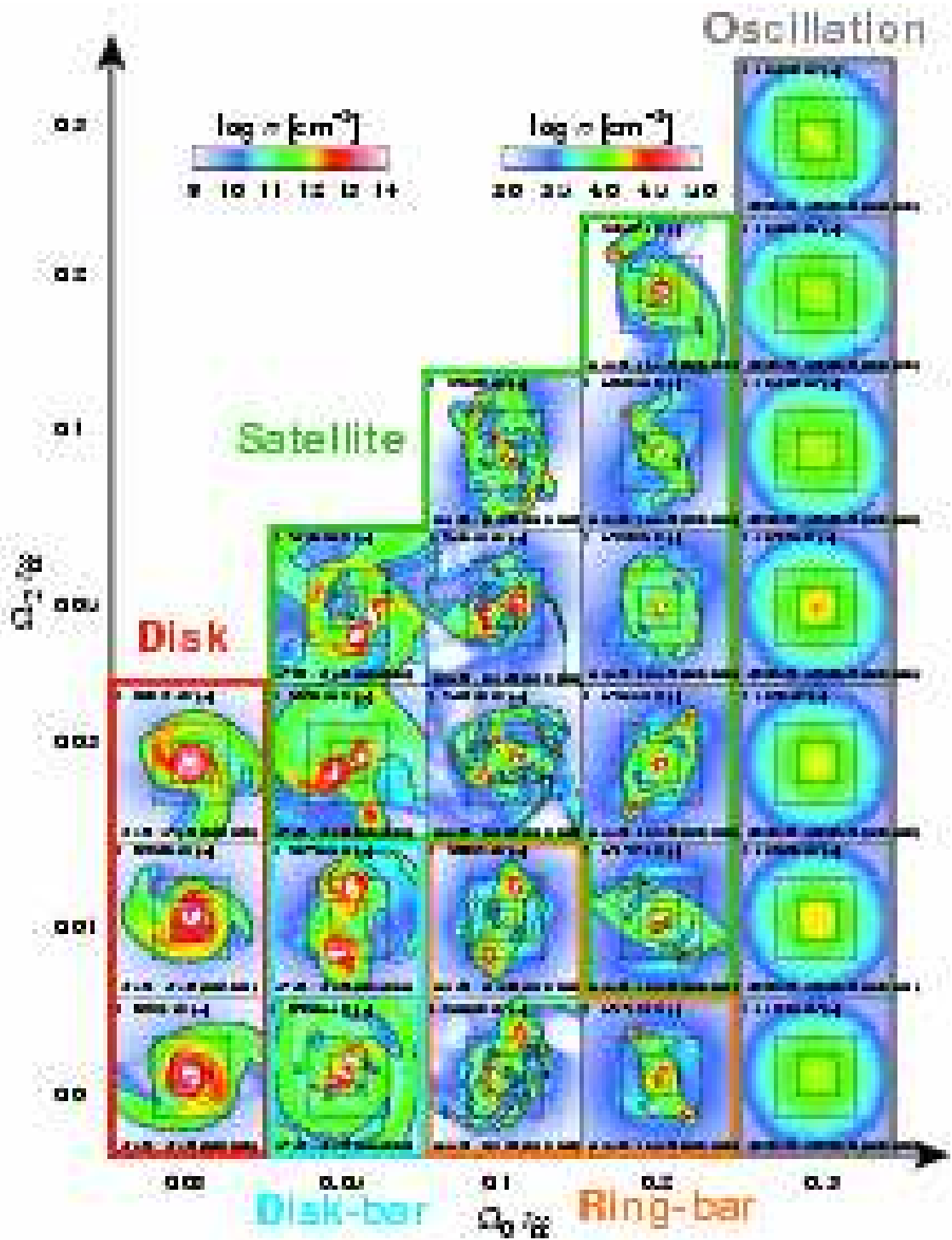} 
\figcaption[rhojc_BE1.1c0.eps]{ 
Density distributions in the $z=0$ plane
at the last stages for models of $C=0$.  Color denotes the density
distribution on a logarithmic scale.  The right color scale is for
{\it oscillation} models, and the left color scale is for models of {\it
disk}, {\it satellite}, {\it ring-bar}, and {\it disk-bar} types.  Black
contour curves denote the critical density $n_{\rm cr}$.  Panels
are arranged in the order of $\Omega_0 t_{\rm ff} = 0.03$, 0.05, 0.1,
0.2, 0.3 from left to right, and $\Omega_2 t_{\rm ff} = 0.0$, 0.01,
0.03, 0.05, 0.1, 0.2, 0.3 from bottom to top.
\label{rhojc_BE1.1c0}
}
\end{figure}

Figure \ref{rhojc_BE1.1c0} summarizes the last stages of the 27 models.
Each panel denotes the density distribution in the $ z = 0 $ plane.
The panels are arranged such that  $\Omega _0$
increases from left to right and  $ \Omega _2 $ increases
from bottom to top. 
The evolutions of the clouds are classified into five types in the
parameter space of the initial rotation ($\Omega_0 t_{\rm ff}$) and the initial
amplitude of bar mode ($\Omega_2 t_{\rm ff}$). 
When $\Omega_0 t_{\rm ff} \le 0.03 $ (left column), 
the cloud collapses to form a single disk ({\it disk} type collapse).
When $\Omega_0 t_{\rm ff} = 0.3$ (right column), 
the cloud never collapses, and instead oscillates. 
When $0.05 \le \Omega_0 t_{\rm ff} \le 0.2$ (middle three columns), 
the cloud collapses into several fragments. 
The last type is further subdivided into three types: 
{\it disk-bar},
{\it ring-bar},
and {\it satellite} types.
In the following, each type is discussed by showing a typical model.

\subsubsection{Disk type collapse}
\label{sec:disk}
\begin{figure}
\epsscale{0.9}
\plotone{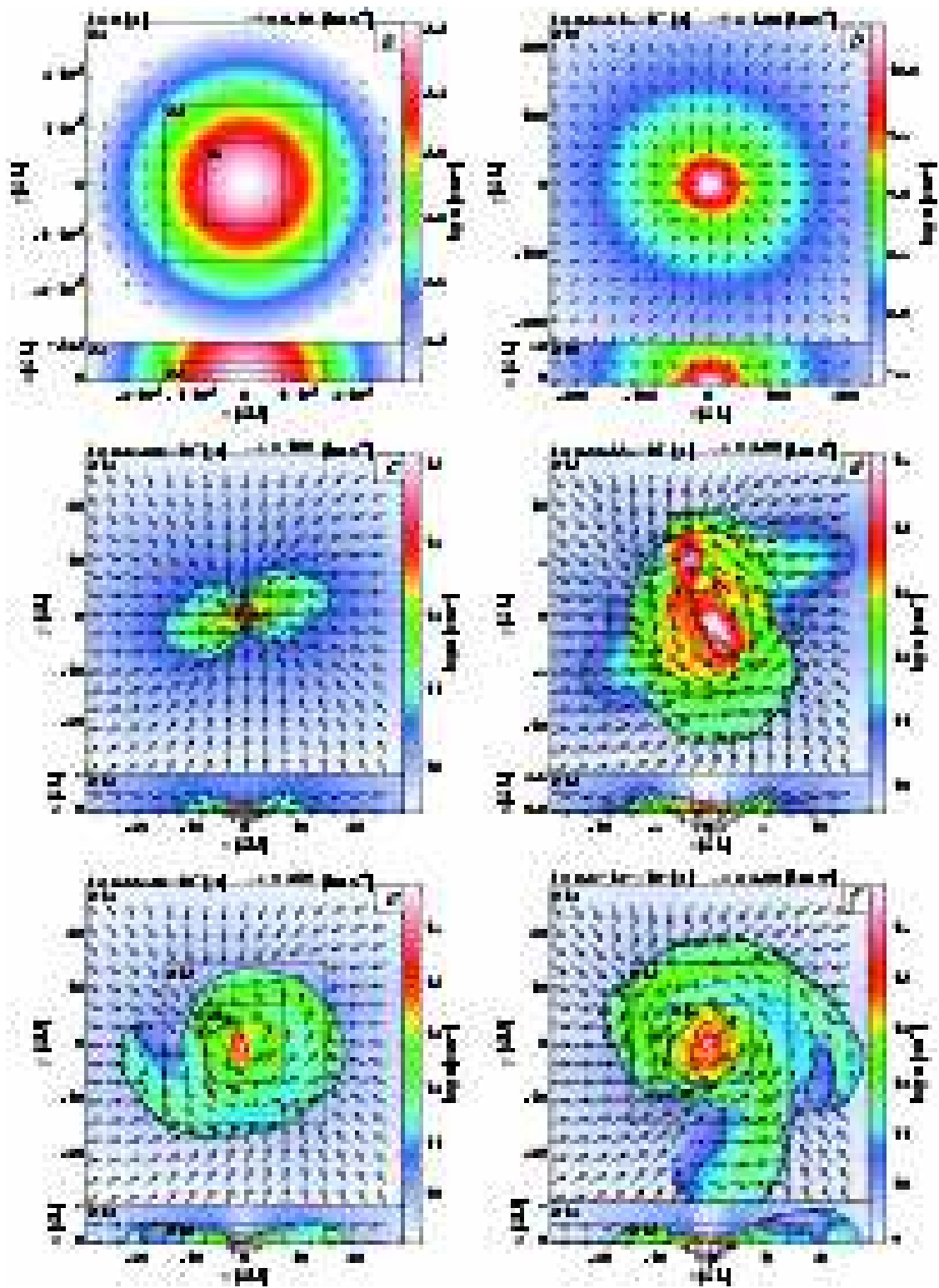}
\figcaption[BE1.1c0_3.e-2_3.e-2_1.e-3.eps]{
Density and velocity distributions for a model of {\it disk} type collapse,
($\Omega_0 t_{\rm ff}$, $\Omega_2 t_{\rm ff}$, $C$) = (0.03, 0.03, 0.0).
Upper and lower panels show the cross sections in the $z=0$ and $y=0$
planes, respectively.
Color scale denotes the density
distribution on a logarithmic scale.
Contour curves denote the critical density $n_{\rm cr}$.
Arrows denote the velocity.
\label{BE1.1c0_3.e-2_3.e-2_1.e-3.eps}
}
\end{figure}

In {\it disk} type collapse, the cloud collapses almost spherically in the
isothermal collapse phase and forms a rotating disk after the central
density exceeds the critical density $\rho_{\rm cr}$.  The disk grows
by accretion and exhibits no sign of fragmentation. 
The model of ($\Omega t_{\rm ff}, \Omega_2 t_{\rm ff}$, $C$) = (0.03, 0.03, 0.0)
is shown in Figure \ref{BE1.1c0_3.e-2_3.e-2_1.e-3.eps}
as typical model of {\it disk} type collapse.

Figure \ref{BE1.1c0_3.e-2_3.e-2_1.e-3.eps}{\it a} shows
the initial stage, showing only the finest three grids
($ 3 \leq l \leq 5 $), i.e., only 1/64 of the full computation
volume.  
The initial
cloud has a spherical density distribution, and the cloud collapses almost
spherically during the isothermal collapse phase ($ \rho < \rho_{\rm cr} $)
as a result of the very slow rotation.
When $\rho \simeq \rho_{\rm cr}$, the central cloud becomes slightly
flattened by the rotation and deforms non-axisymmetrically
due to perturbation of the bar mode (Figure \ref{BE1.1c0_3.e-2_3.e-2_1.e-3.eps}{\it b}). 

The deformation is evaluated by measuring the moment of inertia,
\begin{equation}
I_{ij} = \int_{\rho \ge 0.1 \rho_{\rm max}} 
(r_i - r_{g,i}) (r_j - r_{g,j}) \rho(\mbox{\boldmath $r$})
d\mbox{\boldmath $r$}\;,
\label{eq:I}
\end{equation}
and the total mass,
\begin{equation}
M = \int_{\rho \ge 0.1 \rho_{\rm max}} 
\rho(\mbox{\boldmath $r$})
d\mbox{\boldmath $r$}\;,
\label{eq:M}
\end{equation}
for the gas of $ \rho \geq 0.1 \rho _{\rm max} $.
The subscripts $ i $ and $ j $ are coordinate labels,
i.e., $ x = r _1 $, $ y = r _2 $, and $ z = r _3 $.
The barycenter is defined as
\begin{equation}
r _{g,i} = \frac{1}{M} \int_{\rho \ge 0.1 \rho_{\rm max}} \rho r_i
d\mbox{\boldmath $r$}\;. 
\label{eq:rg}
\end{equation}
The long axis ($ a _l $), short axis ($ a _s $), and 
length along the $ z $-axis ($a_z$) 
are defined by
\begin{equation}
\left(
\begin{array}{l}
a_l^2 \\ a_s^2 \\ a_z ^2
\end{array}
\right) = \frac{1}{2 M}
\left\{
\begin{array}{l}
I_{11} + I_{22} + \left[ (I_{11}-I_{22})^2+ 4I_{12}^2 \right]^{1/2} \\
I_{11} + I_{22} - \left[ (I_{11}-I_{22})^2+ 4I_{12}^2 \right]^{1/2} \\
2 I_{33}
\end{array}
\right\}\;.
\end{equation}

\begin{figure}
\epsscale{0.9}
\plotone{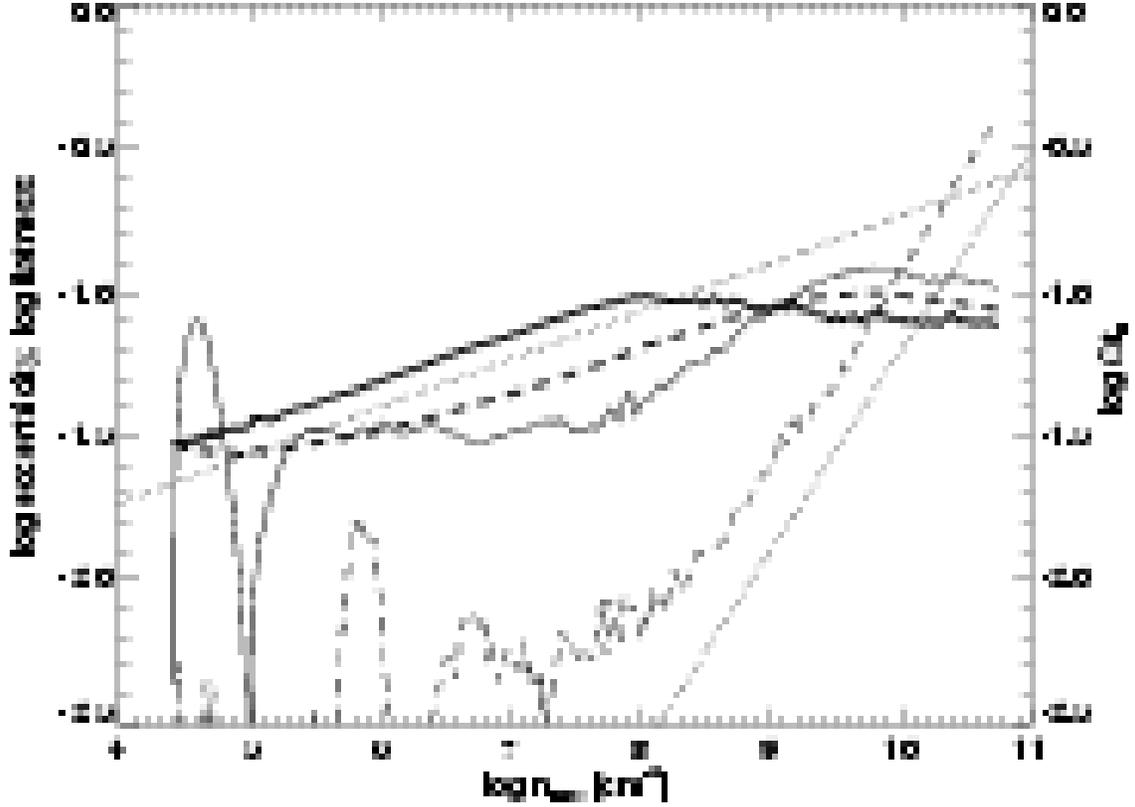}
\figcaption[barlength_plot0.eps]{
Eccentricity, flatness, and angular velocity of the dense region in
the isothermal collapse phase as a function of maximum density for a
model of {\it disk} type collapse, ($\Omega_0 t_{\rm ff}$, $\Omega_2 t_{\rm ff}$, $C$) = (0.03,
0.03, 0.0).  Thin solid and dashed curves denote eccentricity,
$a_l/a_s-1$, and flatness, $(a_l a_s)^{1/2}/a_z -1 $, respectively.
Thick solid and thick dashed curves denote 
angular velocities in a unit of freefall time,
$\Omega_{0.5} t_{\rm ff}$ and $\Omega_{0.1} t_{\rm ff}$, respectively.  
Dotted curves denote the relationships
$\propto n_{\rm max}^{1/6}$ and $n_{\rm max}^{0.7}$ for comparison.
\label{barlength_plot0.eps}
}
\end{figure}

Figure \ref{barlength_plot0.eps} shows 
eccentricity ($a_l/a_s-1$) and flatness [$(a_l a_s)^{1/2}/a_z - 1$] 
as functions of the maximum number density $n_{\rm max}$. 
The eccentricity 
oscillates with significant amplitude in the range
$10^4 \, {\rm cm}^{-3} \leq n_{\rm max}  \lesssim 10^{6}\,
{\rm cm}^{-3}$, and increases roughly in proportion to $n_{\rm
max}^{1/6}$ in the range $10^{6} \, {\rm cm}^{-3} \lesssim n_{\rm
max} \lesssim 10^{10}\,{\rm cm}^{-3}$ due to bar mode
instability \citep{HM99,MH99}.
At the end of the isothermal collapse phase, 
the long axis is 12\% longer than the short axis (eccentricity is 0.122).
The flatness, $a_l/a_z-1$, increases due to spin-up in the central cloud.
In the range $10^{8} \, {\rm cm}^{-3} \lesssim n_{\rm
max}  \lesssim 10^{11}\,{\rm cm}^{-3}$, the flatness increases 
rapidly in proportion to  $n_{\rm max}^{0.7}$. 
At the end of the isothermal collapse phase, the flatness is 0.412 and
the long axis is 50\% longer than thickness in the $z$-direction. 

Figure \ref{barlength_plot0.eps} also shows the central angular
velocity in unit freefall time $\Omega t_{\rm ff}$. 
The angular velocity and the freefall time are measured as
$\Omega = J_{{\rm spin},z}/[M (a_l^2 + a_s^2)]$
and 
$t_{\rm ff} = (3 \pi / 32 G \rho_{\rm max})^{1/2}$,
where $J_{{\rm spin},z}$ denotes the $z$-component of 
the total spin angular momentum $\mbox{\boldmath $J$}_{\rm spin}$,
which is defined as
\begin{equation}
\mbox{\boldmath $J$}_{\rm spin} = \int_{\rho \ge 0.1 \rho_{\rm max}}
(\mbox{\boldmath $r$}-\mbox{\boldmath $r$}_g)
\times
(\mbox{\boldmath $v$}-\mbox{\boldmath $v$}_g)
\rho 
d\mbox{\boldmath $r$}\;,
\end{equation}
where
\begin{equation}
\mbox{\boldmath $v$}_g = \frac{1}{M} \int_{\rho \ge 0.1 \rho_{\rm max}}
\mbox{\boldmath $v$} \rho d \mbox{\boldmath $r$}\;.
\end{equation}
This angular velocity represents the average angular
velocity in the region of $\rho \geq 0.1 \rho_{\rm max}$, and is
denoted $\Omega_{0.1}$. Another average angular velocity
$\Omega_{0.5}$, defined in the region of 
$\rho \geq 0.5 \rho_{\rm max}$, is also introduced 
in order to measure the value near the center.
At the initial stage, $\Omega_{0.1}$ and $\Omega_{0.5}$ have the same
value because the initial cloud rotates rigidly. 
In the early isothermal collapse phase,
the central part rotates faster ($\Omega_{0.5} > \Omega_{0.1}$).
For $n_{\rm max} \lesssim 10^{8}\,{\rm cm}^{-3}$, 
the cloud spins up according to $\Omega_{0.5} t_{\rm ff} \propto n_{\rm
max}^{1/6}$, in other words,
 $\Omega_{0.5} \propto n_{\rm max}^{2/3}$.
This spin-up rate coincides with
that expected for spherical collapse 
\citep{HanawaNakayama97}.
The cloud collapses almost spherically in the isothermal 
collapse phase, and significant deformation only occurs near
the end of the isothermal collapse phase. 
The maximum value of $\Omega_{0.5} t_{\rm ff}$ is 0.0973. 

When the central density exceeds $\rho_{\rm cr}$, 
the infall decelerates near the center and an
adiabatic core, the first core, forms.   The first
core of \citet{larson69} is quasi-static and spherical, whereas this
adiabatic core is rotating and disk-like.  The 
adiabatic core formation ends the isothermal collapse phase.
Infall still continues in the region far from the center, and the
adiabatic core accretes gas from the envelope. 
Thus, the period after adiabatic core formation is called
the accretion phase.    
Figure \ref{BE1.1c0_3.e-2_3.e-2_1.e-3.eps}{\it c} shows the adiabatic
core at the stage of $t-t_{\rm cr} = 216 \, {\rm yr}$, where $t_{\rm
cr}$ denotes the time at which $\rho_{\rm max}$ exceeds $\rho_{\rm cr}$.
The adiabatic core
consists of a flattened central kernel with an envelope of adiabatic
gas. The kernel has a radius of 2~AU and a thickness of 1.5~AU. 
The mass of the kernel is $M_{13} = 7.4 \times 10^{-3} \, M_\odot$,
where $M_{N}$ denotes the mass measured for $n \geq
10^{N}\,{\rm cm}^{-3}$.

Figure \ref{BE1.1c0_3.e-2_3.e-2_1.e-3.eps}{\it d}
shows the adiabatic core at the stage of $t-t_{\rm cr} = 775 \,{\rm yr}$, 
after it has begun to accrete gas from the isothermal infalling envelope.
The adiabatic core at this stage consists of a flattened central kernel
with an extended disk and spiral arms.  One of the dense
spiral arms evolves into a dense clump, as can be seen to the upper left of the
central kernel in Figure
\ref{BE1.1c0_3.e-2_3.e-2_1.e-3.eps}{\it d}. 
The dense clump falls into the central kernel and merges into it.
Figure \ref{BE1.1c0_3.e-2_3.e-2_1.e-3.eps}{\it e} shows the adiabatic
core after the merge of the dense clump ($t-t_{\rm cr} = 838 \,{\rm yr}$).  
The radius of the adiabatic disk increases to 15~AU by this stage. 
During the period
$1.0 \times 10^3 \,{\rm yr} \lesssim t-t_{\rm cr} \lesssim 1.1\times10^{3}\,{\rm yr}$,
this formation and merging of a clump occurs again.
During the formation of the clump, the disk shrinks to 10~AU.
However after merging, the radius of the adiabatic disk increases to 20~AU.  
Figure \ref{BE1.1c0_3.e-2_3.e-2_1.e-3.eps}{\it f} shows the last stage
of the simulation
($t-t_{\rm cr} = 1.5 \times 10^3 \,{\rm yr}$).
The radius of the adiabatic disk has increased to 20~AU, although the radius 
of the central kernel remains.  
The masses of the adiabatic core and the kernel are $6.9\times10^{-2}\,M_\odot$ and
$4.2 \times 10^{-2} \,M_\odot$ at this stage.

\begin{figure}
\epsscale{0.9}
\plotone{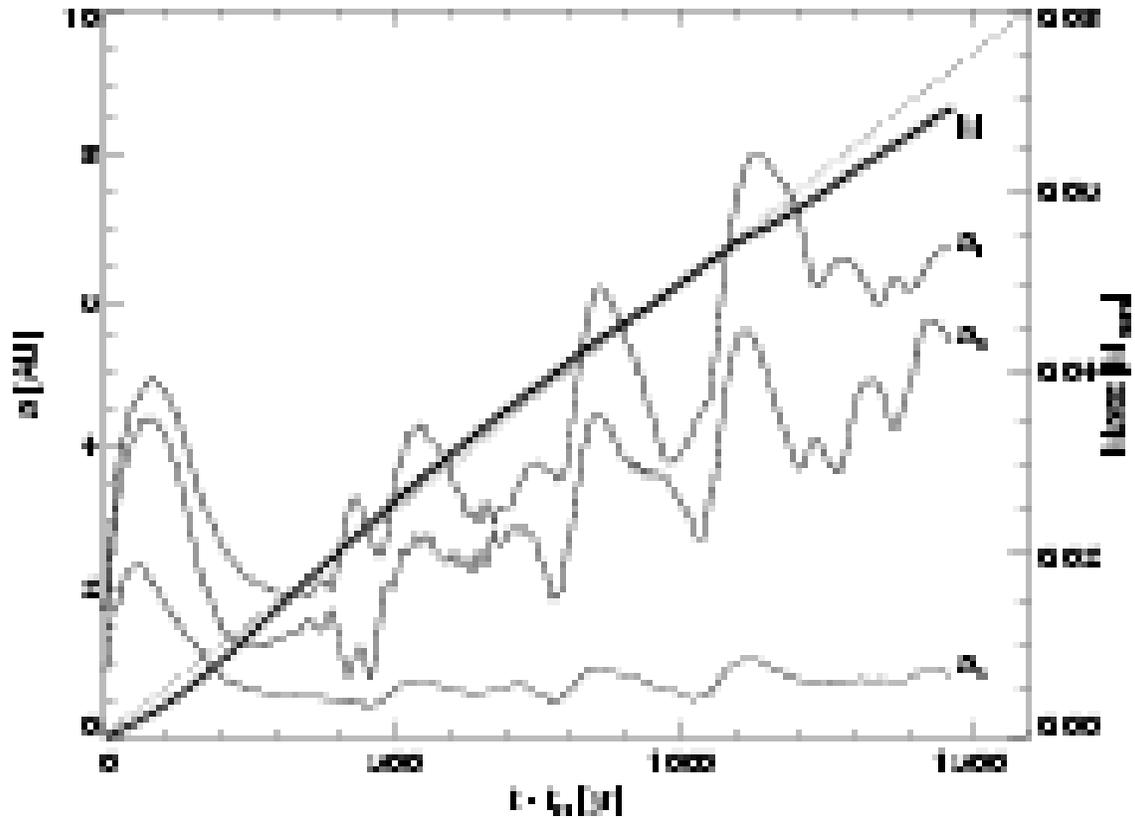}
\figcaption[barlength_plot.eps]{
Size of the adiabatic core and mass in the accretion phase
as a function of time for
a model of {\it disk} type collapse, ($\Omega_0 t_{\rm ff}$, $\Omega_2 t_{\rm ff}$, $C$) = (0.03,
0.03, 0.0).
The three thin curves denote $a_l$, $a_s$, and $a_z$,
the thick curve denotes the mass, and the dotted line denotes 
$M= 5\times 10 ^{-5} (t-t_{\rm cr}) \,M_\odot\,{\rm yr}^{-1}$ for comparison.
\label{barlength_plot.eps}
}
\end{figure}

Figure \ref{barlength_plot.eps} shows the sizes of the adiabatic core
$(a_l, a_s, a_z)$ as a function of time.
These sizes were evaluated in the
same manner as in Figure \ref{barlength_plot0.eps} except
that the volume of integration was bounded by $\rho \ge \rho_{\rm cr}$ in
equations (\ref{eq:I}), (\ref{eq:M}), and (\ref{eq:rg}). 
Note that the sizes evaluated from the moment of inertia correspond to
the scale length of the cloud and are several times smaller than the cloud
radius.  For example, $ a _l $, $ a _s $, and $ a _z $ are 
by a factor of $ 5^{-1/2} = 0.447 $ smaller 
than the radius $a$ for a spherical cloud
having uniform density.
The vertical scale height remains approximately $a_z = 1 - 1.5$~AU,
whereas $a_l$ and $a_s$ increase while oscillating significantly.
The average growth rate is roughly $da_l/dt = da_s/dt\sim3\times10^{-3}
\,{\rm AU}\,{\rm yr}^{-1}$.
These sizes undergo two types of oscillation beyond the steady growth.  First,
large peaks appear near $t - t_{\rm cr} = 800$~yr and 1100~yr.  This
synchronous oscillation corresponds to the formation and merging of the
dense clumps in the adiabatic disk.  The disk has a small radius
during formation, and then expands greatly after merging.  Second, $a_l$ and
$a_s$ exhibit an anti-correlation with small amplitude on a short
timescale. This anti-correlation is due to intermittent excitation of 
spiral arms,  which transfer angular momentum from the
adiabatic disk to the outer infalling envelope.  
These recurrently excited spiral arms are also seen in \citet*{SHM02},
in which growth of the first core was investigated.

Figure \ref{barlength_plot.eps} also shows the mass of the adiabatic core as a
function of time.  The mass increases monotonically due to accretion.
The average accretion rate is approximately 
$5 \times 10^{-5}\,M_\odot$ yr$^{-1}$.

\begin{figure}
\epsscale{0.5}
\plotone{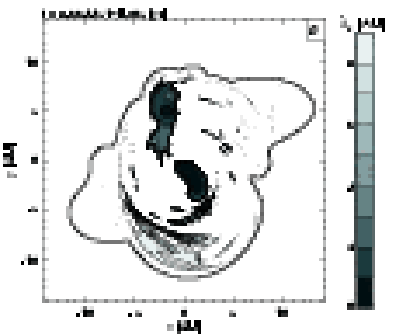}
\plotone{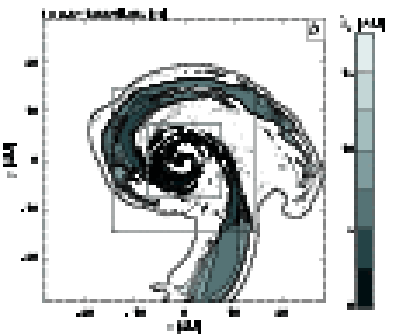}
\figcaption[lambdac_BE1.1c0_3.e-2_3.e-2_1.e-3.eps]{
Distribution of the critical Jeans length $\lambda_c$ for a model of
{\it disk} type collapse, 
($\Omega_0 t_{\rm ff}$, $\Omega_2 t_{\rm ff}$, $C$) = (0.03, 0.03, 0.0). 
Thick contour curve denotes the outline of the adiabatic disk.
The grayscale denotes the critical Jeans length $\lambda_c$.
$Q>1$ in the white region.
Figures ({\it a}) and ({\it b}) correspond to the stages of 
Figures
\ref{BE1.1c0_3.e-2_3.e-2_1.e-3.eps}{\it d} and 
\ref{BE1.1c0_3.e-2_3.e-2_1.e-3.eps}{\it f}, respectively. 
\label{lambdac_BE1.1c0_3.e-2_3.e-2_1.e-3.eps}
}
\end{figure}

The formation of the dense clump and 
excitation of the spiral arms are 
examined here based on the linear stability of \citet{Toomre64}.
A thin disk is unstable when the following two conditions
are satisfied simultaneously.  First, the Toomre $ Q $-value
defined as
\begin{equation}
Q = \frac{c \kappa}{\pi G \Sigma}
\end{equation}
must be smaller than unity in the region of interest, where
$ c $, $  \kappa $ and $ \Sigma $ are the sound speed,
the epicyclic frequency,
and the surface density, respectively.
Second, the unstable region should be larger than
the critical Jeans length,
\begin{equation}
\lambda_c = \frac{2c^2}{G\Sigma}
\left[1+\left(1-Q^2\right)^{1/2}\right]^{-1} \; .
\end{equation}
These criteria are applied to our simulation by 
evaluating 
\begin{equation}
\Sigma(x, y) = \int_{\rho > \rho_{\rm cr}} \rho(x, y, z) dz\;,
\end{equation}
\begin{equation}
c(x, y) = \frac{1}{\Sigma(x,y)}\int_{\rho > \rho_{\rm cr}} \left[
\frac{dP}{d\rho}(x,y,z) \right]^{1/2} \rho(x,y,z) dz \;,
\end{equation}
\begin{equation}
\kappa(x,y) = \left( 4 \bar{\Omega}^2 + R \frac{d\bar{\Omega}^2}{dR}\right)^{1/2}\;,
\end{equation}
and
\begin{equation}
\bar{\Omega}(x,y) = \frac{1}{\Sigma(x,y)}\int_{\rho > \rho_{\rm cr}}
\frac{v_\varphi (x,y,z) }{ R } \rho(x,y,z) dz \;.
\end{equation}
Figure \ref{lambdac_BE1.1c0_3.e-2_3.e-2_1.e-3.eps}{\it a} shows the
distribution of the critical Jeans length $\lambda_c(x,y)$ 
for the stage shown in Figure \ref{BE1.1c0_3.e-2_3.e-2_1.e-3.eps}{\it d}.
The surface density $\Sigma$ takes a positive value 
within the domain indicated by the closed thick curve, i.e.,
edge of the adiabatic disk. 
The Toomre $Q$-value is less than unity in the gray regions, and
larger than unity in the white regions.
It is less than unity only in the central kernel, dense clump and
spiral arms. 
The rest of the disk has $ Q > 1 $ and the disk is
globally stable against the ring mode.
The critical Jeans lengths are less than 1~AU in both the central kernel
and the dense clump, which is roughly 1~AU in size.  
Thus, both are self-gravitationally bounded. 
Figure \ref{lambdac_BE1.1c0_3.e-2_3.e-2_1.e-3.eps}{\it b} shows the
same results for the last stage.  
The $Q$-value is less than unity only in the kernel and the spiral arms.
The spiral arms have width of 2.5~AU in the central
region of $ r \lesssim 5\,{\rm AU} $, where the
critical Jeans length is less than 2.5~AU.
In the outer region of $r \gtrsim 10\,{\rm AU}$, 
the critical Jeans length is 
$5 - 7.5$~AU and the spiral arms have
width of  7.5~AU. 
Thus, the inner and outer spiral arms are self-gravitating
and can be supposed to form by gravitational instability.

\subsubsection{Disk-bar type fragmentation}
\label{sec:disk-bar}

\begin{figure}
\epsscale{0.9}
\plotone{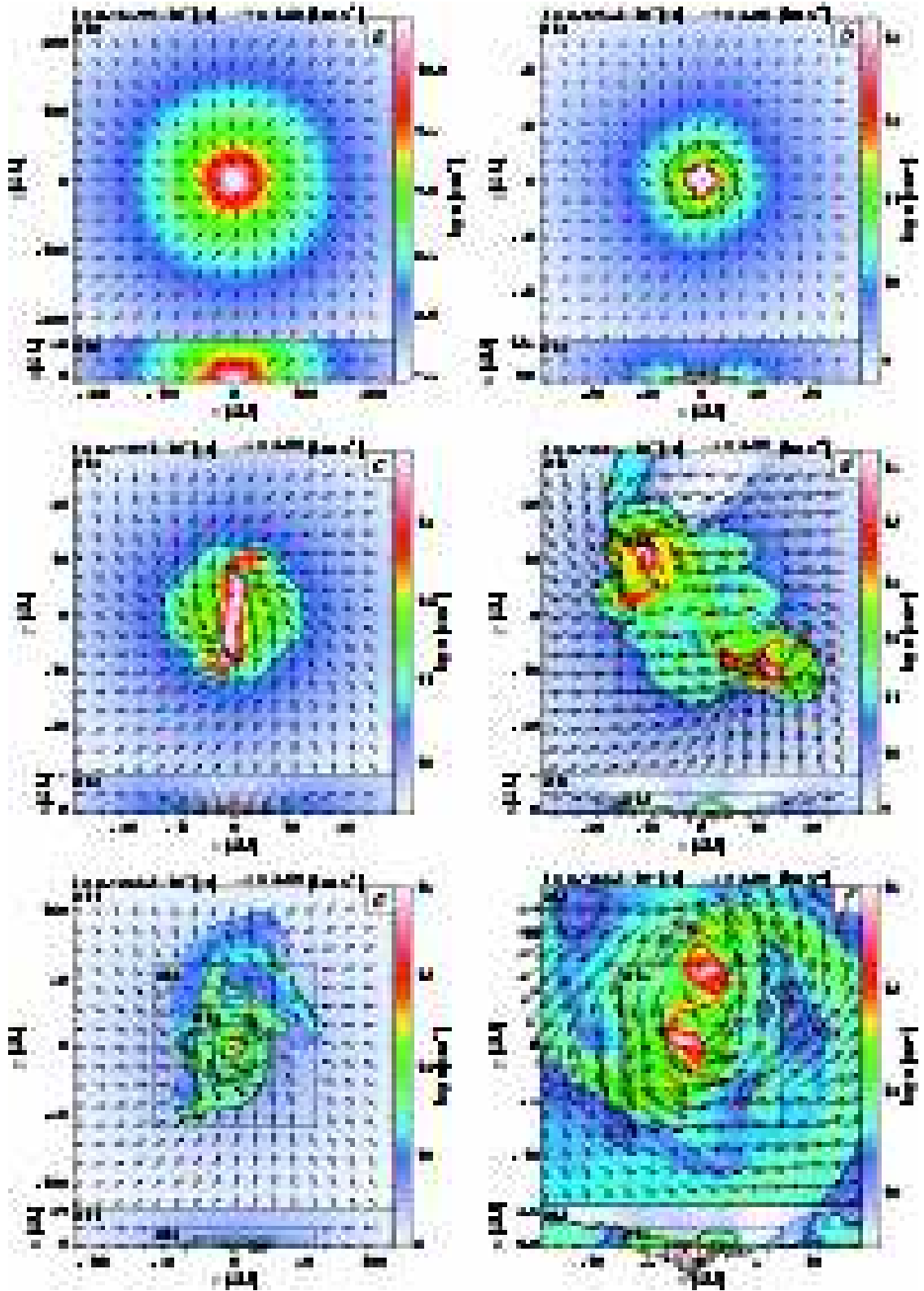}
\figcaption[BE1.1c0_5.e-2_0.e+0_1.e-3.eps]{
Same as Figure \ref{BE1.1c0_3.e-2_3.e-2_1.e-3.eps} but for a model of
{\it disk-bar} type fragmentation, 
($\Omega_0 t_{\rm ff}$, $\Omega_2 t_{\rm ff}$, $C$) = (0.05, 0.0, 0.0). 
Figure \ref{BE1.1c0_5.e-2_0.e+0_1.e-3.eps}{\it f} is an enlargement of 
Figure \ref{BE1.1c0_5.e-2_0.e+0_1.e-3.eps}{\it e}.
\label{BE1.1c0_5.e-2_0.e+0_1.e-3.eps}
}
\end{figure}

In the models of {\it disk-bar} type fragmentation,
$(\Omega_0 t_{\rm ff},\,\Omega_2 t_{\rm
ff},\,C)=(0.05,\,0.0,\,0.0)$ and $(0.05,\,0.01,\,0.0)$, 
the cloud collapse is almost axisymmetric in the isothermal collapse
phase resulting in the formation of a disk 
in the beginning of the accretion phase.
The disk deforms to a bar-shape in the early accretion phase.
Thereafter, the bar fragments into two fragments.
This fragmentation is called {\it disk-bar} type.
The model of 
($\Omega_0 t_{\rm ff}$, $\Omega_2 t_{\rm ff}$, $C$) = (0.05, 0.0, 0.0) 
is shown
in Figure \ref{BE1.1c0_5.e-2_0.e+0_1.e-3.eps}
as a typical model of {\it disk-bar} type fragmentation.

In the initial stage, 
the cloud undergoes uniform
rotation with a small $m=3$ perturbation (no $m=2$ perturbation).
The cloud collapse is almost axisymmetric in the isothermal collapse phase. 
Figure \ref{BE1.1c0_5.e-2_0.e+0_1.e-3.eps}{\it b} shows the cloud at the
end of the isothermal collapse phase. 
The central cloud is slightly flattened due to the rotation. 

\begin{figure}
\epsscale{0.9}
\plotone{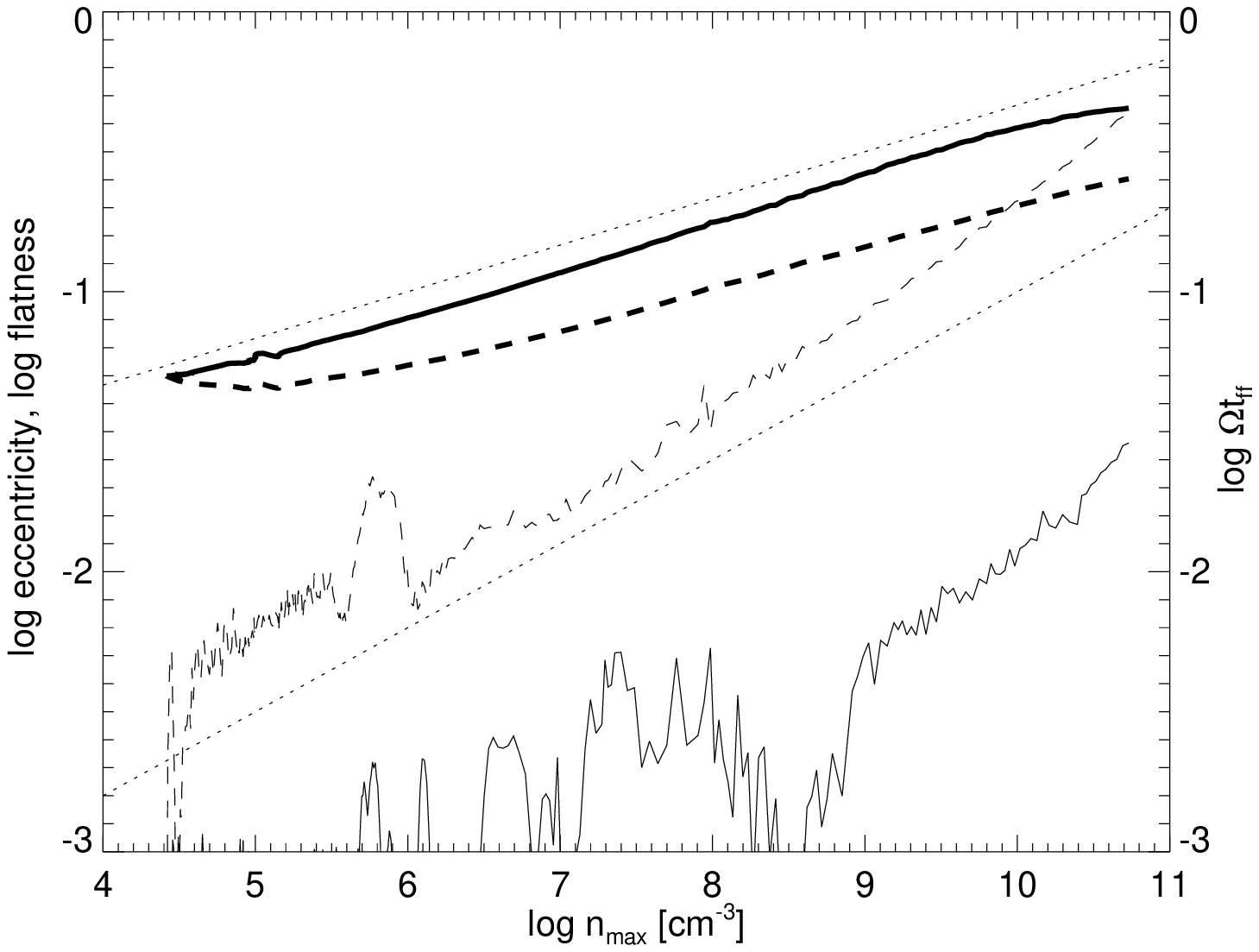}
\figcaption[barlength_plot0_rb.eps]{
Same as Figure \ref{barlength_plot0.eps} but for a model of
{\it disk-bar} type fragmentation, 
($\Omega_0 t_{\rm ff}$, $\Omega_2 t_{\rm ff}$, $C$) = (0.05, 0.0, 0.0).
Dotted curves denote the relationships $\propto n_{\rm max}^{1/6}$ and
$n_{\rm max}^{0.3}$ for comparison.
\label{barlength_plot0_db.eps}
}
\end{figure}

Figure \ref{barlength_plot0_db.eps} shows the evolution of 
flatness [$(a_l a_s)^{1/2}/a_z - 1$], eccentricity ($a_l/a_s-1$), and
angular velocity ($\Omega t_{\rm ff}$)
for the central cloud in the isothermal collapse phase.
The flatness increases rapidly in proportion to $\rho_{\rm
max}^{0.3}$. 
At the end of the isothermal collapse phase, a disk forms with
flatness of 0.439.
The eccentricity remains small
because the initial cloud has only a small $m=3$ perturbation,
with no bar mode perturbation. 
The angular velocity ($\Omega t_{\rm ff}$) is relatively
large at the beginning, and increases according to $\Omega_{0.5} \propto n_{\rm
max}^{1/6}$
up to $ \Omega_{0.5} t _{\rm ff} = 0.452 $ in the isothermal
collapse phase.
The growth rate of $ \Omega_{0.5} t _{\rm ff} $ is 
similar to that shown in Figure \ref{barlength_plot0.eps} for the 
{\it disk} type collapse model.

\begin{figure}
\epsscale{0.9}
\plotone{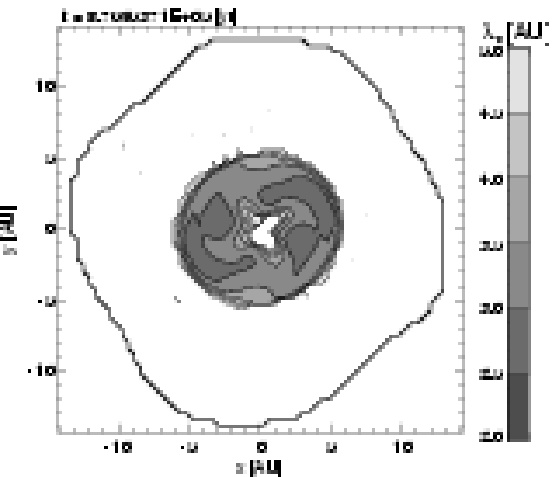}
\figcaption[lambdac_BE1.1c0_5.e-2_0.e+0_1.e-3.eps]{
Same as Figure \ref{lambdac_BE1.1c0_3.e-2_3.e-2_1.e-3.eps} but for
a model of 
{\it disk-bar} type fragmentation, 
($\Omega_0 t_{\rm ff}$, $\Omega_2 t_{\rm ff}$, $C$) = (0.05,
0.0, 0.0)
at the same stage as in Figure \ref{BE1.1c0_5.e-2_0.e+0_1.e-3.eps}{\it b}.  
\label{lambdac_BE1.1c0_5.e-2_0.e+0_1.e-3.eps}
}
\end{figure}

Figure \ref{BE1.1c0_5.e-2_0.e+0_1.e-3.eps}{\it b} shows the adiabatic
core at the stage of $t-t_{\rm cr} = 301$~yr.
The adiabatic core at this time consists of a flat disk with an
envelope.  The adiabatic disk deforms to a bar-shape at 
$t-t_{\rm cr} = 439$~yr, as shown in 
Figure \ref{BE1.1c0_5.e-2_0.e+0_1.e-3.eps}{\it c},
due to self-gravitational instability.  
The seed of the bar mode is discretization error in this model.
Figure \ref{lambdac_BE1.1c0_5.e-2_0.e+0_1.e-3.eps} shows the distribution
of critical Jeans length for the stage shown in Figure
\ref{BE1.1c0_5.e-2_0.e+0_1.e-3.eps}{\it b}.  
The elliptical disk of 12~AU $\times$ 11~AU has a critical Jeans
length of $2-4$~AU, except in the region of the central holes, $r \lesssim
2 \,{\rm AU}$.  The disk thus suffers from gravitational
instability and deforms to a bar. 
The bar fragments into two fragments at $t-t_{\rm cr}\simeq 500$~yr,
with masses of $M_{13} = 1.0 \times 10^{-2} \, M_\odot$ and
$1.1 \times 10^{-2} \, M_\odot$, and 
$M_{14} = 4.8 \times 10^{-3} \, M_\odot$ and $3.0 \times 10^{-3} \, M_\odot$.
The fragments rotate around each other and accrete gas
from the envelope. 
Figure \ref{BE1.1c0_5.e-2_0.e+0_1.e-3.eps}{\it d} shows the 
fragments at the stage of $t-t_{\rm cr} = 1622$~yr. 
Similar to the adiabatic core shown in Figure
\ref{BE1.1c0_3.e-2_3.e-2_1.e-3.eps}{\it f}, 
each fragment
has a central kernel and spiral arms embedded in an extended disk.
At this stage, the two fragments have mass of
$M_{13}=3.0 \times 10^{-2} \, M_\odot $ and $ 2.4 \times 10^{-2} \, M_\odot$, and 
$M_{14}=1.8 \times 10^{-2} \, M_\odot $ and $ 1.9 \times 10^{-2} \, M_\odot$.

\begin{figure}
\epsscale{0.7}
\plotone{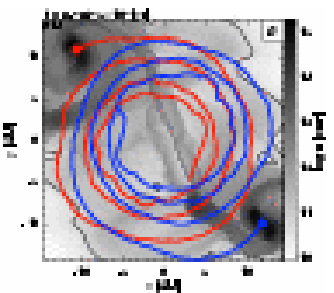}
\plotone{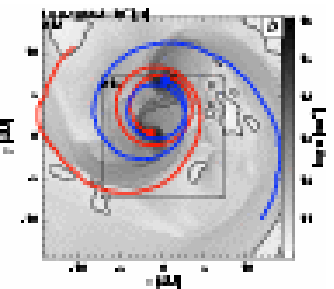}
\figcaption[BE1.1c0_5.e-2_0.e+0_1.e-3_orbit.eps]{
Loci of fragments in the $x-y$ plane for the model of
{\it disk-bar} type fragmentation, 
($\Omega_0 t_{\rm ff}$, $\Omega_2 t_{\rm ff}$, $C$) = (0.05, 0.0, 0.0). 
Color curves trace the barycenters of each fragment 
($n \geq 10^{14}\,{\rm cm}^{-3}$) 
in the period ({\it a}) between the stage of fragmentation 
and maximum separation, and ({\it b}) in the rest stages.
The grayscales in figures ({\it a}) and ({\it b}) denote
the density distributions at the stages of Figures
\ref{BE1.1c0_5.e-2_0.e+0_1.e-3.eps}{\it d} and
\ref{BE1.1c0_5.e-2_0.e+0_1.e-3.eps}{\it f}, respectively. 
\label{BE1.1c0_5.e-2_0.e+0_1.e-3_orbit.eps}
}
\end{figure}

\begin{figure}
\epsscale{0.9}
\plotone{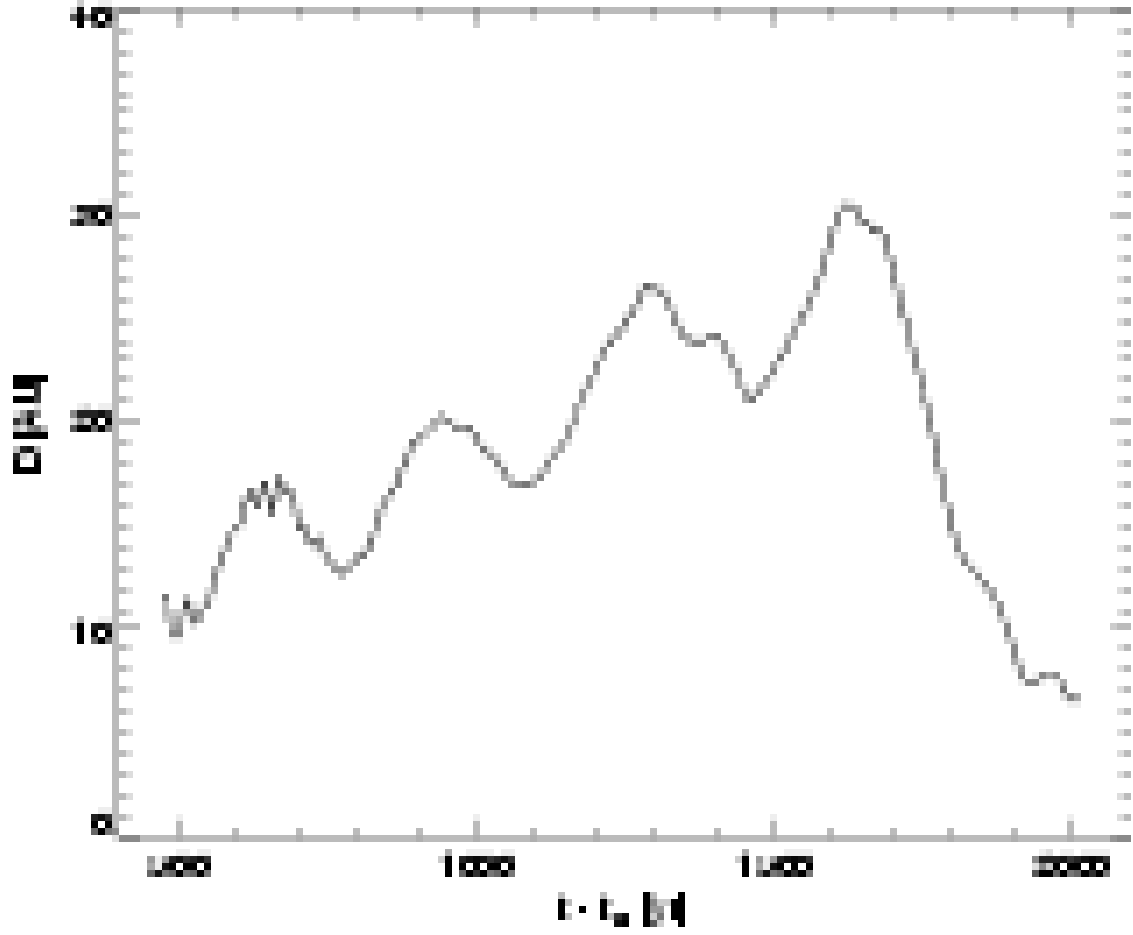}
\figcaption[clumpdist]{
Separation between fragments as a function of time for 
{\it disk-bar} type model,
($\Omega_0 t_{\rm ff}$, $\Omega_2 t_{\rm ff}$, $C$) = (0.05, 0.0, 0.0).
\label{clumpdist.eps}
}
\end{figure}

Figure \ref{BE1.1c0_5.e-2_0.e+0_1.e-3_orbit.eps}{\it a} shows
the loci of fragments for the period between the stages
of fragmentation and Figure \ref{BE1.1c0_5.e-2_0.e+0_1.e-3.eps}{\it d}
(468~yr  $ \leq t - t_{\rm cr} \leq $ 1622~yr).
In this period of $1.1 \times 10^3$~yr, 
both fragments rotate approximately three and half times,
increasing in separation.
The time variation of the separation is shown quantitatively in Figure
\ref{clumpdist.eps}.
The separation of the fragments
increases from 11.6~AU with significant oscillation due to the eccentricity of the orbits.
The fragments attain a maximum separation of 30.6~AU at $t-t_{\rm cr} = 1622$~yr 
(at the stage of Figure \ref{BE1.1c0_5.e-2_0.e+0_1.e-3.eps}{\it d}).
As the separation increases, the specific orbital angular momentum
also increases by a factor of 4.9. 

Figure \ref{BE1.1c0_5.e-2_0.e+0_1.e-3_orbit.eps}{\it b} shows the situation
after the maximum separation.  After the stage of Figure
\ref{BE1.1c0_5.e-2_0.e+0_1.e-3.eps}{\it d}, the separation begins to
decrease, to $\sim 7$~AU in only $\sim 300 $~yr,
as shown in Figure \ref{clumpdist.eps}.  The specific orbital
angular momentum also decreases by a factor of 0.28.  In this period, the
fragments rotate only half a rotation.  After the rapid decrease in the
separation, the fragments rotate approximately 1.5 revolutions with a
nearly constant separation of $\sim 7$~AU until the last stage.
Figures \ref{BE1.1c0_5.e-2_0.e+0_1.e-3.eps}{\it e} and 
\ref{BE1.1c0_5.e-2_0.e+0_1.e-3.eps}{\it f} show the last stage
of the simulation at different magnifications. 
At this stage, the fragments are separated by 6.7~AU,
and surrounded by a circumbinary disk with tightly winding spiral arms
that transfer the orbital angular momentum of the fragments to the
circumbinary disk.  
The decrease in separation is due to the formation of the circumbinary
disk. 
At the last stage, the two fragments have the same mass of 
$M_{14}=2.4 \times 10^{-2}\, M_\odot$.

\subsubsection{Ring-bar type fragmentation}
\label{sec:ring-bar}

\begin{figure}
\epsscale{0.9}
\plotone{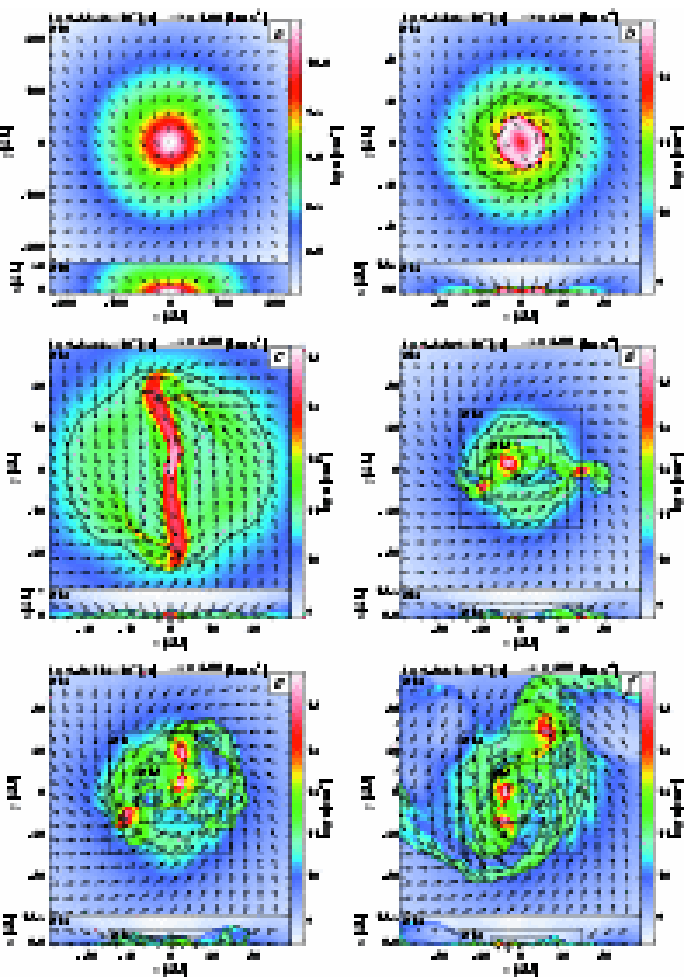}
\figcaption[BE1.1c0_1.e-1_0.e+0_1.e-3.eps]{
Same as Figure \ref{BE1.1c0_3.e-2_3.e-2_1.e-3.eps} but for a model of
{\it ring-bar} type fragmentation,
($\Omega_0 t_{\rm ff}$, $\Omega_2 t_{\rm ff}$, $C$) = (0.1, 0.0, 0.0). 
\label{BE1.1c0_1.e-1_0.e+0_1.e-3.eps}
}
\end{figure}

In the models showing
{\it ring-bar} type fragmentation,
$(\Omega_0 t_{\rm ff},\,\Omega_2 t_{\rm ff},\,C) = 
(0.1,\,0.0,\,0.0)$,
$(0.1,\,0.01,\,0.0)$,
and $(0.2,\,0.0,\,0.0)$, 
cloud collapse is almost axisymmetric and a flat 
disk forms in the isothermal collapse phase. 
The disk deforms to a ring-shape temporarily,
and then to a bar-shape in the early accretion phase.
Thereafter, the bar fragments into two or three fragments.  
This fragmentation is called {\it ring-bar} type fragmentation.
The model of 
($\Omega_0 t_{\rm ff}$, $\Omega_2 t_{\rm ff}$, $C$) = (0.1, 0.0, 0.0) 
is shown 
in Figure \ref{BE1.1c0_1.e-1_0.e+0_1.e-3.eps} 
as a typical example of {\it ring-bar} type fragmentation. 
This type of fragmentation is similar to {\it disk-bar}
except for the transient formation of a ring and the number of fragments.

The initial stage is the same as for the previous model shown in
\S \ref{sec:disk-bar} except for the initial uniform rotation speed.
The cloud collapse is almost axisymmetric in the isothermal collapse phase. 
Figure \ref{BE1.1c0_1.e-1_0.e+0_1.e-3.eps}{\it a} shows the cloud at the
end of the isothermal collapse phase. The central cloud
deforms to a disk-shape due to the rotation. 

\begin{figure}
\epsscale{0.9}
\plotone{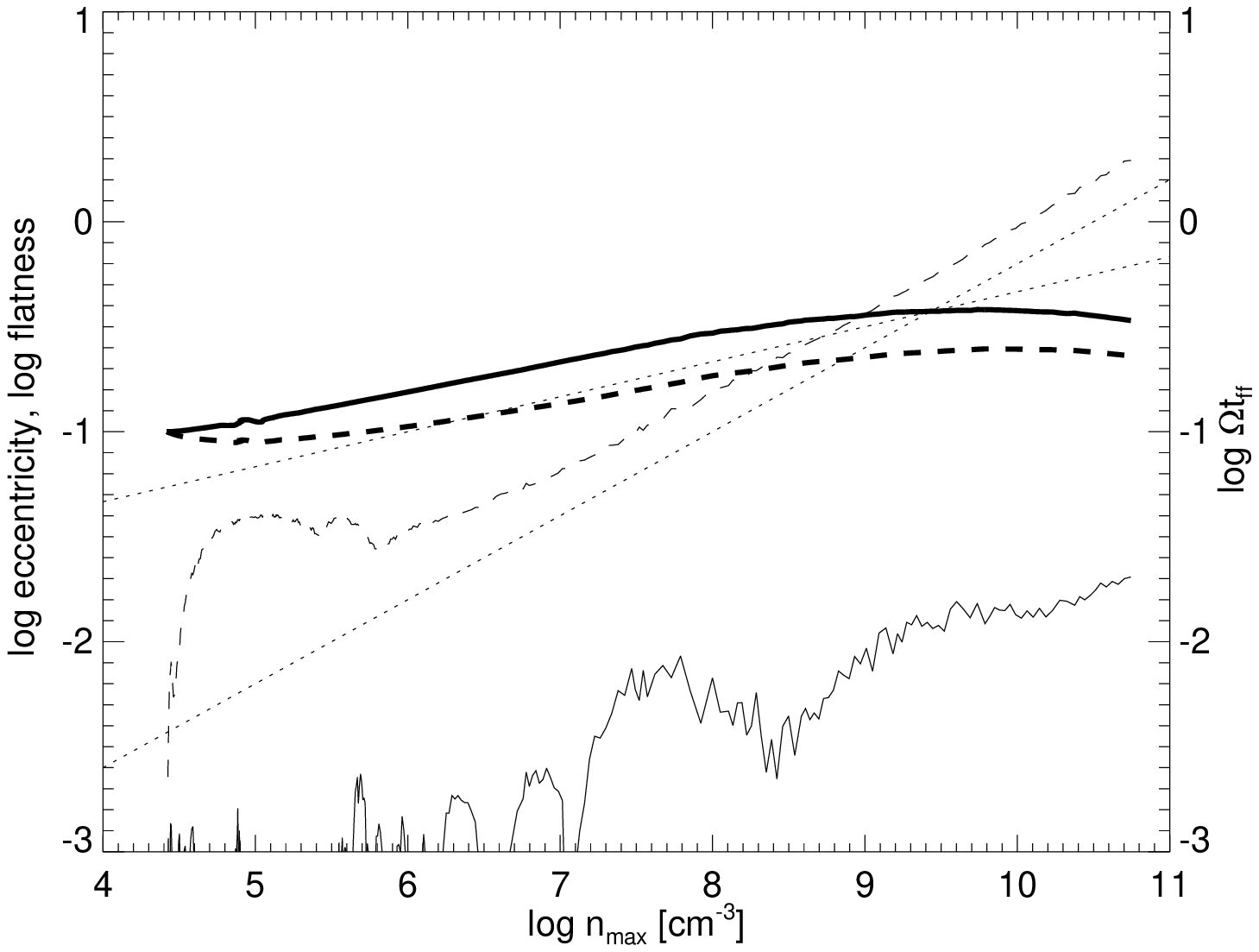}
\figcaption[barlength_plot0_rb.eps]{
Same as Figure \ref{barlength_plot0.eps} but for a model of
{\it ring-bar} type fragmentation,
($\Omega_0 t_{\rm ff}$, $\Omega_2 t_{\rm ff}$, $C$) = (0.1, 0.0, 0.0).
Dotted curves denote the relationships $\propto n_{\rm max}^{1/6}$ and
$n_{\rm max}^{0.4}$ for comparison.
\label{barlength_plot0_rb.eps}
}
\end{figure}

Figure \ref{barlength_plot0_rb.eps} shows the evolution of 
flatness [$(a_l a_s)^{1/2}/a_z - 1$], eccentricity ($a_l/a_s-1$), and
angular velocity ($\Omega t_{\rm ff}$)
for the central cloud in the isothermal collapse phase.
These evolutions are similar to those for the {\it disk-bar} model.
The flatness increases rapidly in proportion to $\rho_{\rm
max}^{0.4}$. 
At the end of the isothermal collapse phase, a thin disk has formed, with
flatness of 1.96. 
The eccentricity remains small
because the initial cloud has only a small $m=3$ perturbation  
and no bar mode perturbation. 
The angular velocity ($\Omega t_{\rm ff}$) is relatively
large from the beginning, and increases according to 
$\Omega_{0.5} t_{\rm ff} \propto n_{\rm max}^{1/6}$
to a maximum of $ \Omega_{0.5} t _{\rm ff} = 0.381 $
at $n_{\rm max} = 5.46 \times 10^9\,{\rm cm}^{-3}$.
When the central cloud is disk-like, 
$ \Omega t _{\rm ff} $ becomes saturated.

\begin{figure}
\epsscale{0.9}
\plotone{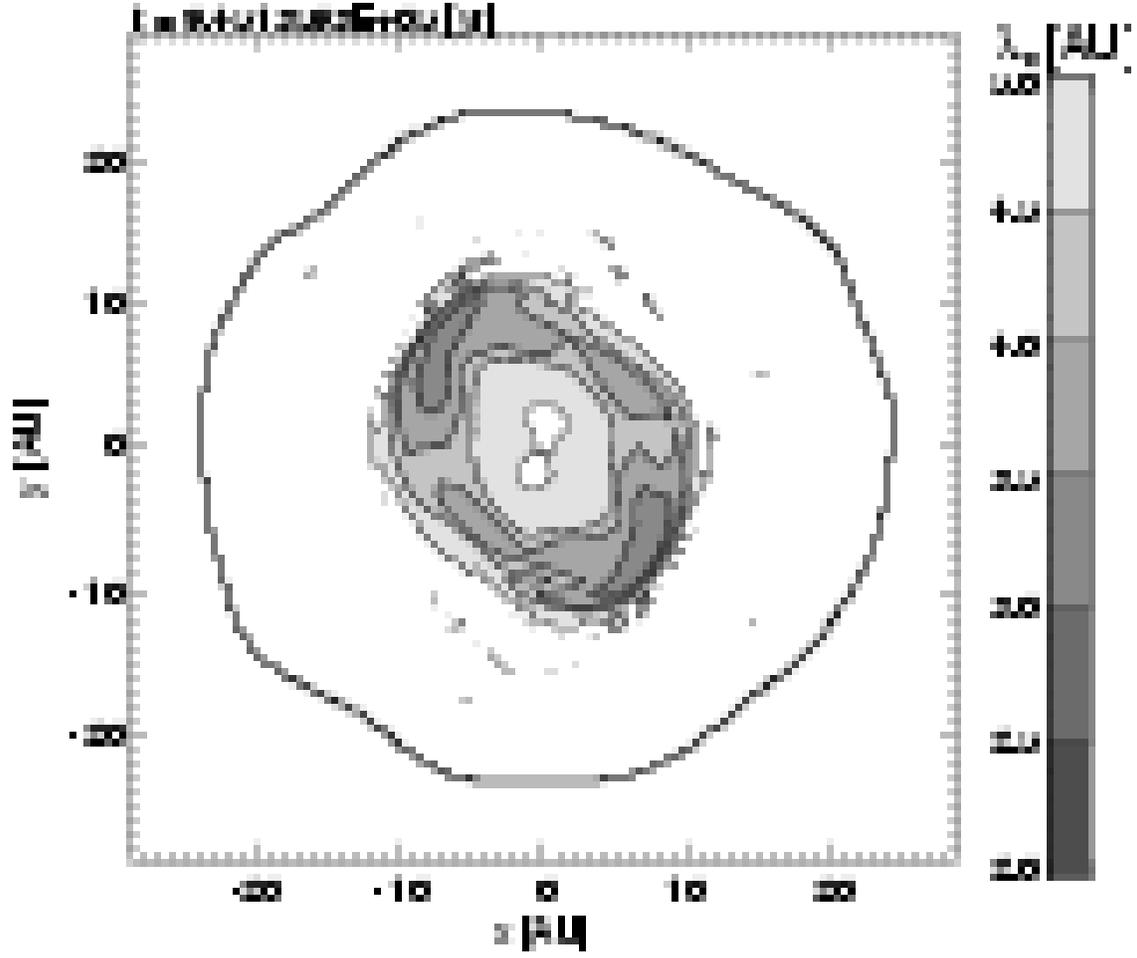}
\figcaption[lambdac_BE1.1c0_1.e-1_0.e+0_1.e-3.eps]{
Same as Figure \ref{lambdac_BE1.1c0_3.e-2_3.e-2_1.e-3.eps} but for
a model of {\it ring-bar} type fragmentation,
($\Omega_0 t_{\rm ff}$, $\Omega_2 t_{\rm ff}$, $C$) = (0.1,
0.0, 0.0)
at the same stage as in Figure \ref{BE1.1c0_1.e-1_0.e+0_1.e-3.eps}{\it b}. 
\label{lambdac_BE1.1c0_1.e-1_0.e+0_1.e-3.eps}
}
\end{figure}

Figure \ref{BE1.1c0_1.e-1_0.e+0_1.e-3.eps}{\it b} shows the adiabatic core
at $t-t_{\rm cr} = 563$~yr.
The central adiabatic core is ring-like 
and surrounded by a flat isothermal envelope.
The ring structure forms due to self-gravitational instability. 
Figure \ref{lambdac_BE1.1c0_1.e-1_0.e+0_1.e-3.eps} shows the distribution
of critical Jeans length for the stage shown in Figure
\ref{BE1.1c0_1.e-1_0.e+0_1.e-3.eps}{\it b}.  
The distribution is similar to that for the {\it
disk-bar} model except that the adiabatic elliptical disk is
larger roughly by a factor of two while the critical Jeans length is
almost the same.
The disk thus suffers from ring instability more strongly than
in the {\it disk-bar} model.

The ring-shaped adiabatic core deforms to a
rotating bar as shown in Figure \ref{BE1.1c0_1.e-1_0.e+0_1.e-3.eps}{\it c}.
The bar is twice as long as that in the {\it disk-bar} model,
and the bar fragments into three fragments. 
Figure \ref{BE1.1c0_1.e-1_0.e+0_1.e-3.eps}{\it d} shows the fragments
at $t-t_{\rm cr} = 915$~yr.
The central fragment is the most massive ($M_{13}= 2.0 \times 10^{-2}\,M_\odot$).
The other fragments have masses of $M_{13}=4.8 \times10^{-3}\,M_\odot$
and $5.4 \times10^{-3}\,M_\odot$.

\begin{figure}
\epsscale{0.9}
\plotone{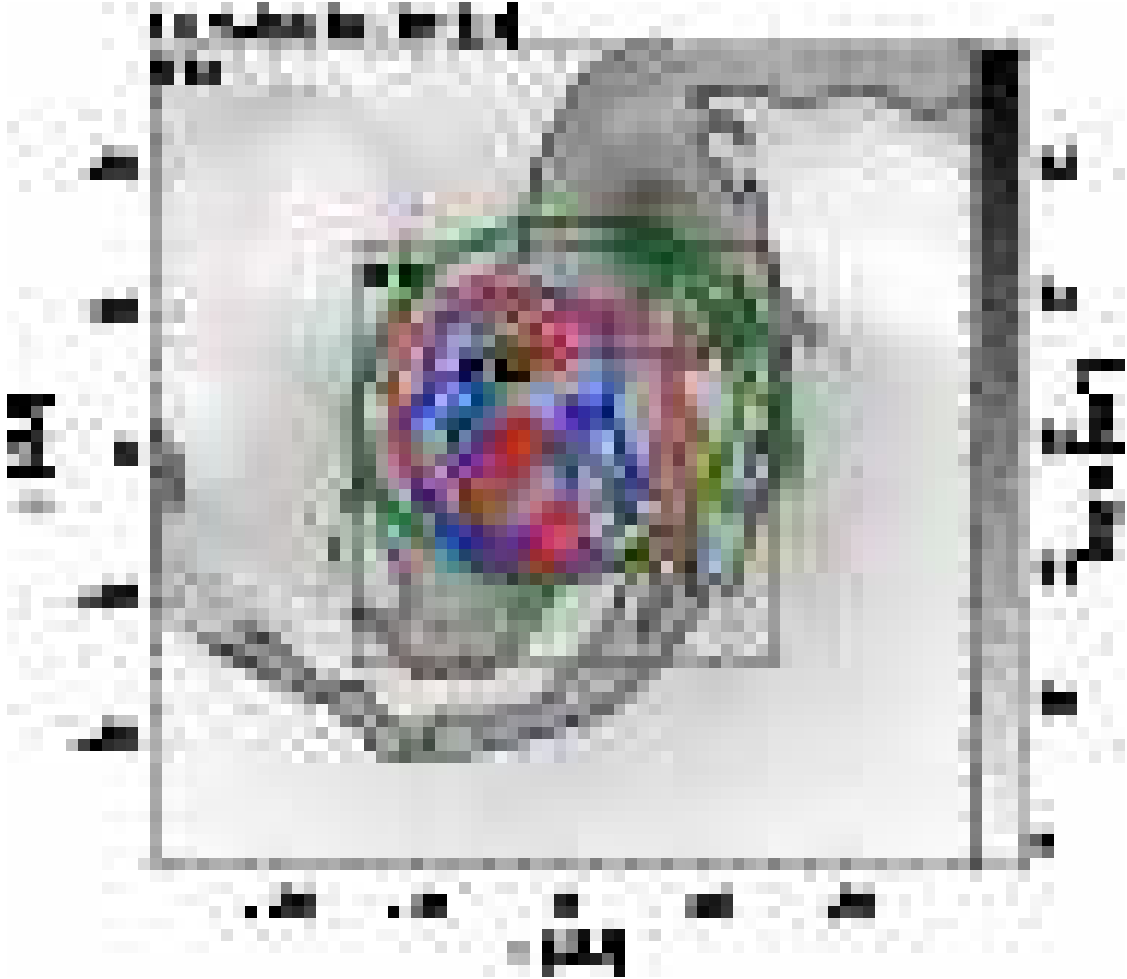}
\figcaption[BE1.1c0_1.e-1_0.e+0_1.e-3_orbit.eps]{
Same as Figure \ref{BE1.1c0_5.e-2_0.e+0_1.e-3_orbit.eps} but for 
a model of {\it ring-bar} type fragmentation,
($\Omega_0 t_{\rm ff}$, $\Omega_2 t_{\rm ff}$, $C$) = (0.1, 0.0, 0.0). 
Red and blue loci trace the barycenters
of each fragment of $n \geq 10^{14}\,{\rm cm}^{-3}$.
Green locus is for the fragment of $n \geq 10^{13}\,{\rm cm}^{-3}$.
Grayscale denotes the logarithmic density at the same
stage as in
Figure \ref{BE1.1c0_1.e-1_0.e+0_1.e-3.eps}{\it f}. 
\label{BE1.1c0_1.e-1_0.e+0_1.e-3_orbit.eps}
}
\end{figure}

Figure \ref{BE1.1c0_1.e-1_0.e+0_1.e-3_orbit.eps} shows the loci of the
three fragments. The red and blue fragments rotate at a distance 
of roughly 15~AU, while the green fragment rotates around the
red-blue close binary at distance of roughly 40~AU (see also 
Figure \ref{BE1.1c0_1.e-1_0.e+0_1.e-3.eps}{\it e}).
The three fragments form a hierarchical triple system.

Figure \ref{BE1.1c0_1.e-1_0.e+0_1.e-3.eps}{\it f} shows the last stage
of the simulation. The three fragments have similar masses of 
$M_{13} = 1.9 \times 10^{-2} M_\odot$ (red),
$M_{13} = 1.7 \times 10^{-2} M_\odot$ (blue), and 
$M_{13} = 1.8 \times 10^{-2} M_\odot$ (green).
The separation between the red and blue fragments is 14~AU, and
that between the their barycenters and the green fragment is 41.6~AU
at the last stage.

The calculation was terminated at this stage because the green fragment
escaped from the region covered by the fine grid ($l = 13$) to that
covered by the coarser grid ($l=12$).

\subsubsection{Satellite type fragmentation}
\label{sec:satellite}
In the models with
$0.05 \leq \Omega_0 t_{\rm ff} \leq 0.2$ and 
with large $\Omega_2 t_{\rm ff}$, 
the cloud collapses to form a dense adiabatic core surrounded by 
an adiabatic disk. The disk suffers from self-gravitational
instability and fragments into dense fragments
orbiting around the central adiabatic core.  
The orbiting fragment is called a  satellite
fragment, and this fragmentation {\it satellite} type fragmentation.
No appreciable difference between {\it satellite} type and {\it disk}
type is seen up to disk formation.
Spiral arms  are excited in the adiabatic
disk and local dense condensations form.  
The masses of these condensations exceed the 
Jeans mass, and evolve into satellite fragments confined by self-gravity.
The satellite fragments orbit around the central core 
and exhibit close encounter, and sometimes merge during the accretion phase.
In following, the model of 
($\Omega_0 t_{\rm ff}$, $\Omega_2 t_{\rm ff}$, $C$) = (0.1, 0.05, 0.0)
is shown as a typical example.

\begin{figure}
\epsscale{0.9}
\plotone{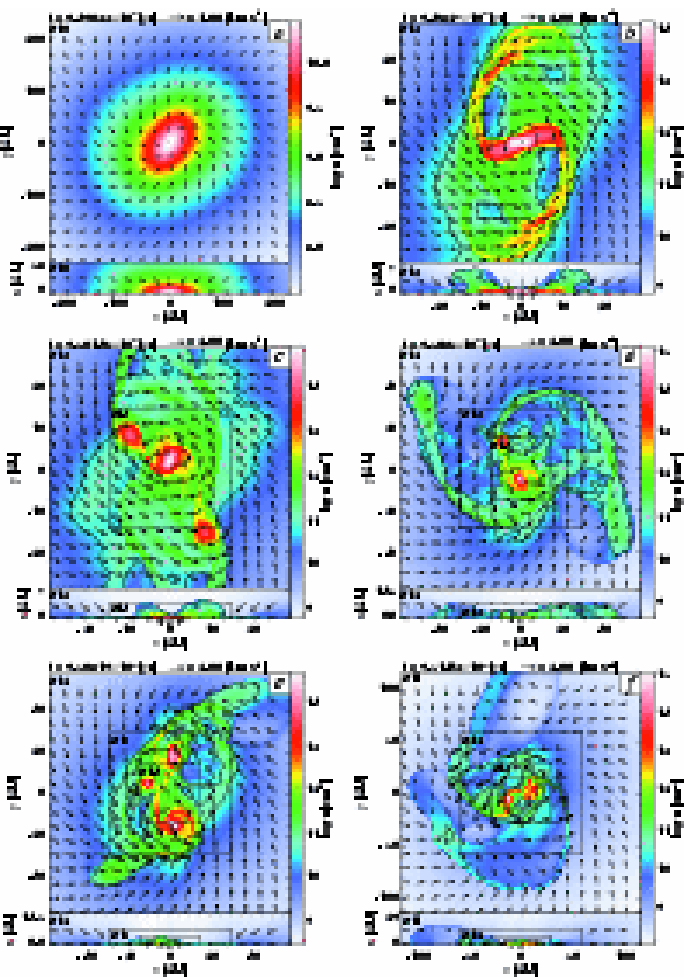}
\figcaption[BE1.1c0_1.e-1_5.e-2_1.e-3.eps]{
Same as Figure \ref{BE1.1c0_3.e-2_3.e-2_1.e-3.eps} but for a model of
{\it satellite} type fragmentation,
($\Omega_0 t_{\rm ff}$, $\Omega_2 t_{\rm ff}$, $C$) = (0.1, 0.05, 0.0). 
\label{BE1.1c0_1.e-1_5.e-2_1.e-3.eps}
}
\end{figure}

For the model of ($\Omega_0 t_{\rm ff}$, $\Omega_2 t_{\rm ff}$, $C$) =
(0.1, 0.05, 0.0), 
the initial stage is the same as that of the previous model ({\it
  ring-bar} type) except for the amplitude of 
the bar mode.  Figure \ref{BE1.1c0_1.e-1_5.e-2_1.e-3.eps}{\it a} shows the
cloud at the end of the isothermal collapse phase.  The dense gas
is flattened due to the rotation, and the cross section in the $y=0$ plane
resembles that of the previous model because of the same initial rotation
(see Figure
\ref{BE1.1c0_1.e-1_0.e+0_1.e-3.eps}{\it a}). The central cloud is
more elongated than in the previous model due to the large initial
amplitude of the bar mode.  

\begin{figure}
\epsscale{0.7}
\plotone{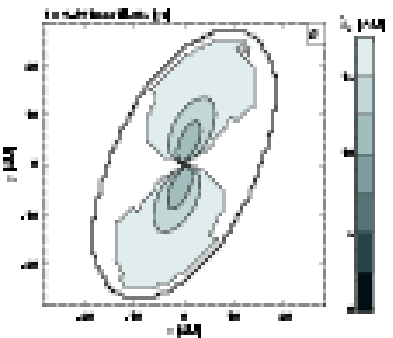}
\plotone{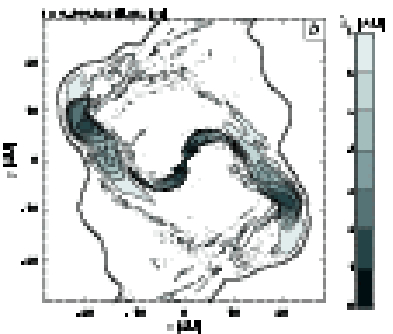}
\figcaption[lambdac_BE1.1c0_1.e-1_5.e-2_1.e-3.eps]{
Same as  Figure \ref{lambdac_BE1.1c0_3.e-2_3.e-2_1.e-3.eps} but for
a model of {\it satellite} type fragmentation,
($\Omega_0 t_{\rm ff}$, $\Omega_2 t_{\rm ff}$, $C$) = (0.1,
0.05, 0.0).
\label{lambdac_BE1.1c0_1.e-1_5.e-2_1.e-3.eps}
}
\end{figure}

Figure \ref{lambdac_BE1.1c0_1.e-1_5.e-2_1.e-3.eps}{\it a} shows the 
distribution of critical Jeans length at the stage of adiabatic bar formation.  
Only a limited region along the axis is unstable against
ring instability in the bar, while the adiabatic core is elongated into a bar. 

Figure \ref{BE1.1c0_1.e-1_5.e-2_1.e-3.eps}{\it b} shows the elongated
adiabatic core.  The central kernel (dense region in the adiabatic core)
rotates differentially and excites spiral arms in 
the surrounding disk.
The elongation of the central kernel precedes excitation of the ring
instability as shown by \citet{SHM02}. 
%
In their simulation, a cloud without the bar mode forms a ring-shaped core,
while a cloud with the bar mode forms an elongated core with spiral
arms.  The amplitude of the bar mode separates the {\it satellite}
type fragmentation from {\it disk-bar} and {\it ring-bar} types. 

The spiral arms sweep up the disk and are wound up.  These winding spiral
arms then evolve into satellite fragments, which are confined by 
self-gravity, as shown in Figure
\ref{BE1.1c0_1.e-1_5.e-2_1.e-3.eps}{\it c}.
Figure \ref{lambdac_BE1.1c0_1.e-1_5.e-2_1.e-3.eps}{\it b} shows the
distribution of critical
Jeans length at the stage of satellite fragment formation. 
The adiabatic disk is globally stable against the ring mode, yet is locally
unstable in small regions around $(x,\,y) \simeq (\pm
20\,{\rm AU},\,\mp 7\,{\rm AU}) $. 
A satellite fragment forms in each of the local unstable regions.

\begin{figure}
\epsscale{0.7}
\plotone{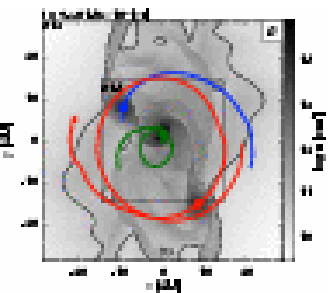}
\plotone{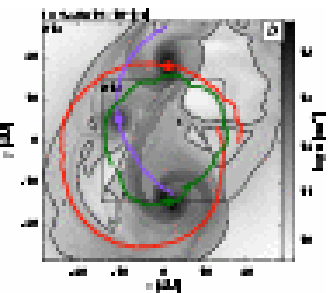}
\figcaption[BE1.1c0_1.e-1_5.e-2_1.e-3_orbit.eps]{
Same as  Figure \ref{BE1.1c0_5.e-2_0.e+0_1.e-3_orbit.eps}
but for a model of {\it satellite} type fragmentation,
($\Omega_0 t_{\rm ff}$, $\Omega_2 t_{\rm ff}$, $C$) =
(0.1, 0.05, 0.0).
Colored curves trace the barycenters of 
({\it a}) each fragment of 
$n \geq 10^{13}\,{\rm cm}^{-3}$ 
in the period between the stages of the first {\it satellite} fragmentation and the next
{\it satellite} fragmentation (formation of the purple fragment), and 
({\it b}) fragments of $n \geq 10^{14}\,{\rm cm}^{-3}$ 
in the remaining period.
Grayscales in figures ({\it a}) and ({\it b})
denote the density distributions at the stages of Figures
\ref{BE1.1c0_1.e-1_5.e-2_1.e-3.eps}{\it c} and
\ref{BE1.1c0_1.e-1_5.e-2_1.e-3.eps}{\it e}, respectively. 
\label{BE1.1c0_1.e-1_5.e-2_1.e-3_orbit.eps}
}
\end{figure}

Figure \ref{BE1.1c0_1.e-1_5.e-2_1.e-3_orbit.eps}{\it a} shows the loci of 
the central fragment and the satellite fragments.
The green locus denotes the central fragment, while the blue and red loci
indicate the satellite fragments.
The blue satellite fragment merges into the central
fragment (green) after half a rotation, and 
the blue and red fragments form a binary system as shown in Figure
\ref{BE1.1c0_1.e-1_5.e-2_1.e-3.eps}{\it d}.
The binary excites spiral arms and a new satellite fragment forms 519~yr
after the merger.
Figure \ref{BE1.1c0_1.e-1_5.e-2_1.e-3_orbit.eps}{\it b} shows 
the loci after formation of the new satellite fragment. 
The purple locus denotes the new satellite fragment.
The smallest fragment in Figure
\ref{BE1.1c0_1.e-1_5.e-2_1.e-3.eps}{\it e} is the new satellite fragment,
which later falls into the green fragment resulting in a 
binary system again.
Figure \ref{BE1.1c0_1.e-1_5.e-2_1.e-3.eps}{\it f} shows the last stage of
the simulation. 
At this stage, the masses of the
fragments are  
$M_{13}= 2.0 \times 10^{-2} M_\odot$ (red) and
$3.0\times 10^{-2} M_\odot$ (green), 
and the separation between fragments is 26.9~AU.

\subsection{Dependence on the Rotation Law}

In this subsection,
the dependence of the collapse and
fragmentation on the rotation law specified by the
parameter $C$ is investigated.
The azimuthally averaged angular velocity is independent of $R$ (uniform)
at $C=0$, and decreases with increasing $R$ as $C$ gets larger.

\begin{figure}
\epsscale{0.9}
\plotone{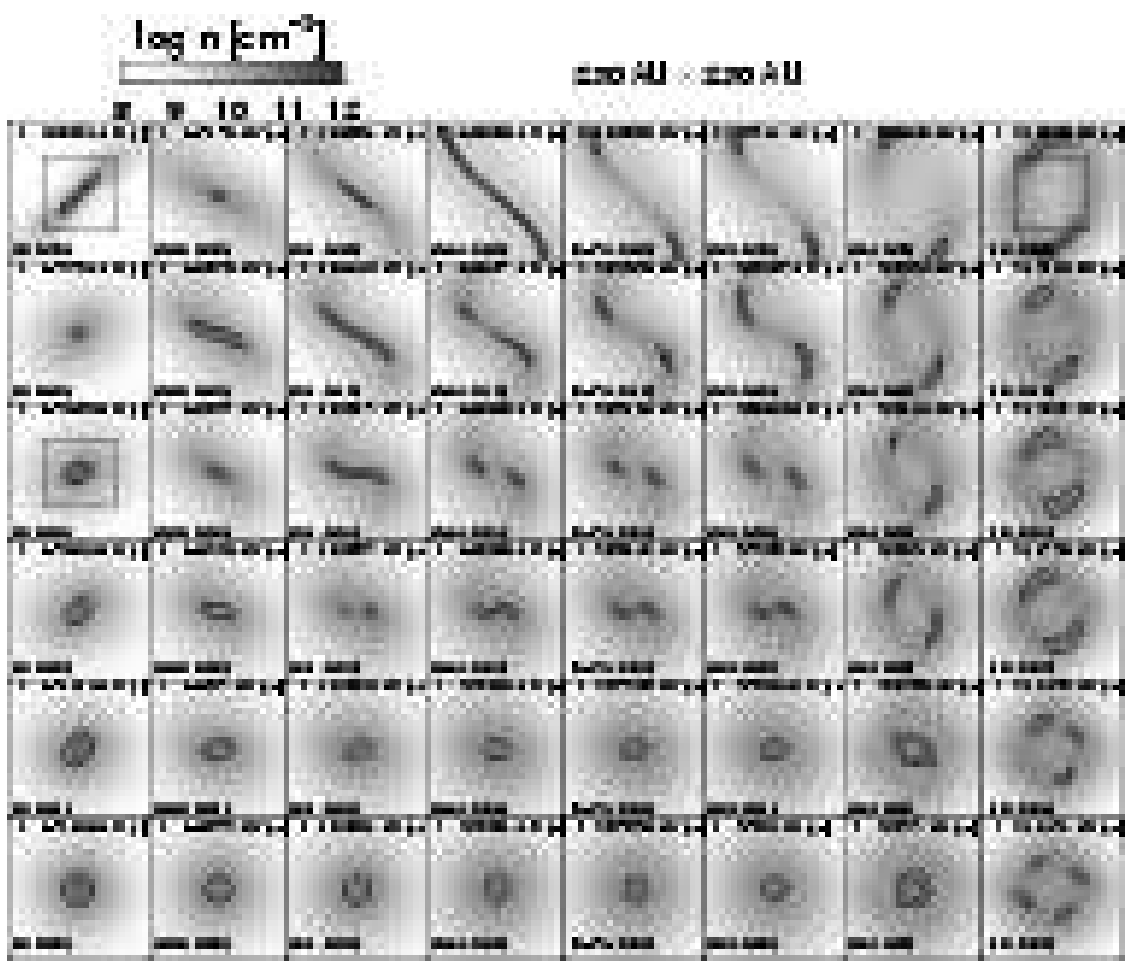}
\figcaption[c-a2.eps]{
Morphology of the central cloud at the stage of $n_{\rm max} \simeq n_{\rm cr}$.
Each panel shows the density distribution 
in the central 230~AU $\times$ 230~AU of the $z=0$ plane.
Contour curves denote the critical density $n_{\rm cr}$.
Parameters ($C$, $\Omega_2 t_{\rm ff}$) are noted at the bottom of
each panel.
Panels are arranged in the order of 
increasing $C$ from left to right,
and 
increasing $\Omega_2 t_{\rm ff}$ from bottom to top.
Initial rotation is $\Omega_0 t_{\rm ff} = 0.2$ for all models.
\label{c-a2.eps}
}
\end{figure}

Figure \ref{c-a2.eps} shows the models of $\Omega_0 t_{\rm ff} = 0.2$
for the central 230~AU $\times$ 230~AU square at the stage of $\rho_c
\simeq \rho_{\rm cr}$.  Based on the morphology, these models can be
classified into three types, i.e., 
{\it ring}, {\it bar}, and {\it dumbbell} types.
Ring-shaped structures are seen in models with large $C$,
while bar-shaped structures occur in models with large $\Omega_2 t_{\rm ff}$.
When both $C$ and $\Omega_2 t_{\rm ff}$ are large, 
the density has two peaks, and forms the {\it dumbbell} type. 
Both the dumbbell and ring forms in the model of  $C \gtrsim 0.16$. 
Fragmentation, particularly the number of fragments, depends critically
on the morphology.

The cloud with $\Omega_0 t_{\rm ff} \lesssim 0.03$ collapses to form
an adiabatic disk and exhibits no sign of
fragmentation ({\it disk} type collapse). 
On the other hand, a cloud with $\Omega_0 t_{\rm ff}
\gtrsim 0.05$ fragments by any of the {\it disk-bar}, {\it ring-bar}, {\it satellite},
{\it ring}, {\it bar}, or {\it dumbbell} types as far as it
collapses.  The parameter $ \Omega _0 t _{\rm ff} $ solely 
specifies whether the cloud fragments, whereas 
the other parameters specify only the type of fragmentation.

\subsubsection{Ring type fragmentation}
\label{sec:ring}

{\it Ring} type fragmentation takes place in models with large $C$
and small $\Omega_2 t_{\rm ff}$.  A ring forms during the collapse and
fragments into more than three fragments.  

\begin{figure}
\epsscale{0.9}
\plotone{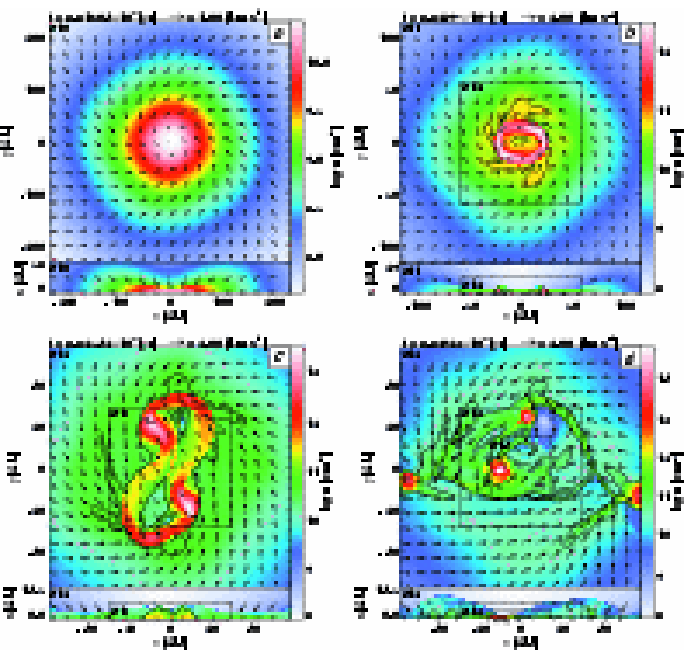}
\figcaption[BE1.1c1.0_2.e-1_0.e+0_1.e-3.eps]{
Same as Figure \ref{BE1.1c0_3.e-2_3.e-2_1.e-3.eps} but for a model of
{\it ring} type fragmentation,
$(\Omega_0 t_{\rm ff},\,\Omega_2 t_{\rm
ff}, \, C)=(0.2,\,0.0,\, 1.0)$.
\label{BE1.1c1.0_2.e-1_0.e+0_1.e-3.eps}
}
\end{figure}

\begin{figure}
\epsscale{0.9}
\plotone{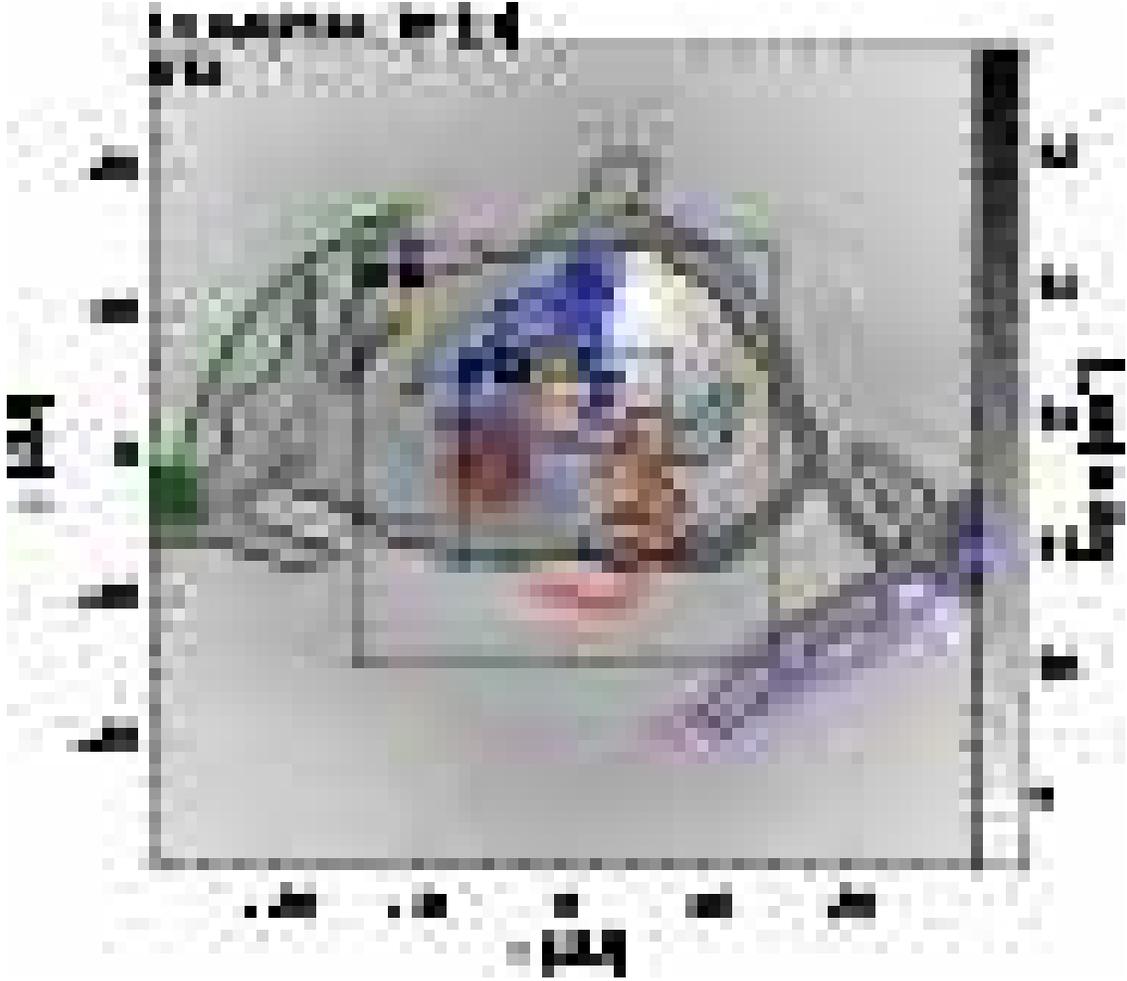}
\figcaption[BE1.1c1.0_2.e-1_0.e+0_1.e-3_orbit.eps]{
Same as  Figure \ref{BE1.1c0_5.e-2_0.e+0_1.e-3_orbit.eps}
but for a model of 
{\it ring} type fragmentation,
($\Omega_0 t_{\rm ff}$, $\Omega_2 t_{\rm ff}$, $C$) =
(0.2, 0.0, 1.0). 
Dashed parts of blue and red loci show schematic orbits,
which can not be defined separately because
the fragments approach too closely.
\label{BE1.1c1.0_2.e-1_0.e+0_1.e-3_orbit.eps}
}
\end{figure}

Figure \ref{BE1.1c1.0_2.e-1_0.e+0_1.e-3.eps} show the model with
$(\Omega_0 t_{\rm ff},\,\Omega_2 t_{\rm ff}, \, C)=(0.2,\,0.0,\, 1.0)$
as a typical model of {\it ring} type fragmentation.  
Figure \ref{BE1.1c1.0_2.e-1_0.e+0_1.e-3.eps}{\it a} shows the central
cloud at the end of the isothermal collapse phase.  
The evolution is similar for {\it ring} type and {\it ring-bar}
type as far as the isothermal collapse phase is concerned,
with a flat disk forming in the cloud center.  
Figure \ref{BE1.1c1.0_2.e-1_0.e+0_1.e-3.eps}{\it b} shows the stage of 
$t-t_{\rm cr}=634$~yr.  The central disk suffers from ring instability,
and the ring is more prominent than for the {\it ring-bar} type.
Figure
\ref{BE1.1c1.0_2.e-1_0.e+0_1.e-3.eps}{\it c} shows 
the stage of $t-t_{\rm cr}=879$~yr, just at the moment of fragmentation.
The ring fragments directly into four fragments, whereas the ring deforms into
a bar before fragmentation in {\it ring-bar} type fragmentation.
Figure
\ref{BE1.1c1.0_2.e-1_0.e+0_1.e-3.eps}{\it d} shows the stage of 
$t-t_{\rm cr}=1.26\times10^{3}$~yr (the last stage), in which four
fragments can be seen. 
The loci of the four fragments are shown in
Figure \ref{BE1.1c1.0_2.e-1_0.e+0_1.e-3_orbit.eps}. 
The two central fragments (blue and red) exhibit close encounter while
the outer fragments (green and purple) rotate with wide orbits.
At the last stage, the masses of the fragments are
$M_{13}= 2.0 \times 10^{-2} M_\odot$ (red),
$ 9.4 \times 10^{-3}\, M_\odot$ (blue), 
$ 7.0 \times 10^{-3}\, M_\odot$ (green), and
$ 7.4 \times 10^{-3}\, M_\odot$ (purple).
The calculation was terminated here because 
the Jeans condition was violated after the escape of the outer fragments
from the fine grid at $ l = 12 $.

\subsubsection{Bar type fragmentation}
\label{sec:bar}

\begin{figure}
\epsscale{0.9}
\plotone{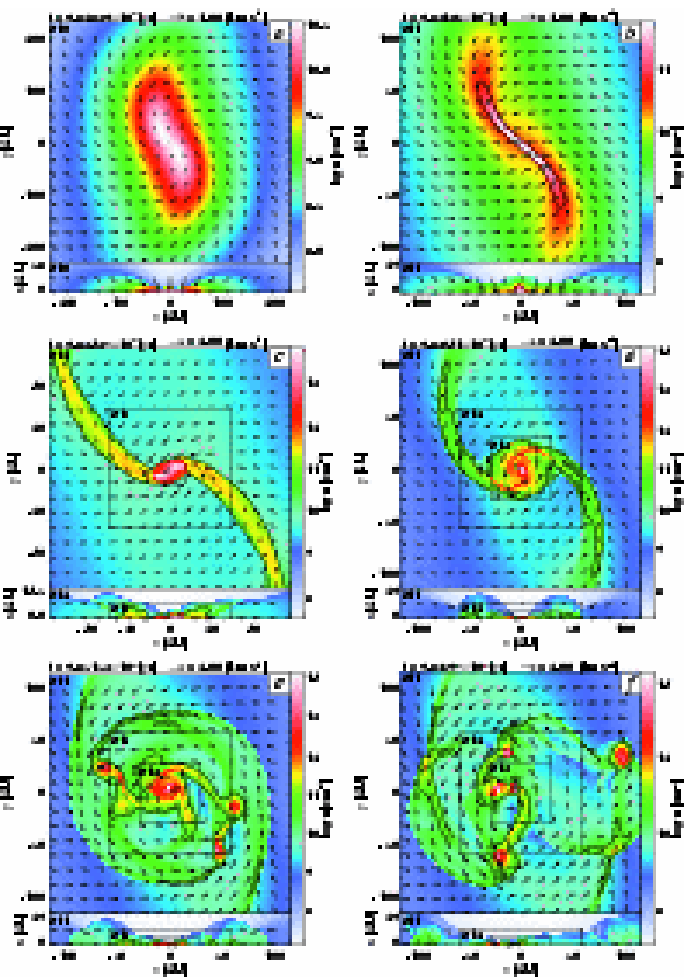}
\figcaption[BE1.1c0.15_2.e-1_2.e-1_1.e-3.eps]{
Same as Figure \ref{BE1.1c0_3.e-2_3.e-2_1.e-3.eps} but for model a of
{\it bar} type fragmentation followed by {\it satellite} type fragmentation,
$(\Omega_0 t_{\rm ff},\,\Omega_2 t_{\rm ff}, \, C)
=(0.2,\,0.2,\, 0.15)$.
\label{BE1.1c0.15_2.e-1_2.e-1_1.e-3.eps}
}
\end{figure}

{\it Bar} type fragmentation takes place in models with 
large $\Omega_2 t_{\rm ff}$.
The cloud collapses to form
a narrow bar. Although the bar fragments into two fragments in many models,
the fragments in this model merge to form a central adiabatic core.
The merger is due to the small angular momentum of the fragments.
After the merger, the adiabatic core excites spiral arms and
eventually {\it satellite} type fragmentation occurs as shown in
\S\ref{sec:satellite}.  
Figure \ref{BE1.1c0.15_2.e-1_2.e-1_1.e-3.eps} shows the model with
$(\Omega_0 t_{\rm ff},\,\Omega_2 t_{\rm ff},\, C) =(0.2,\,0.2,\,
0.15)$ as a typical model of {\it bar} type fragmentation followed by
{\it satellite} type fragmentation.

\begin{figure}
\epsscale{0.9}
\plotone{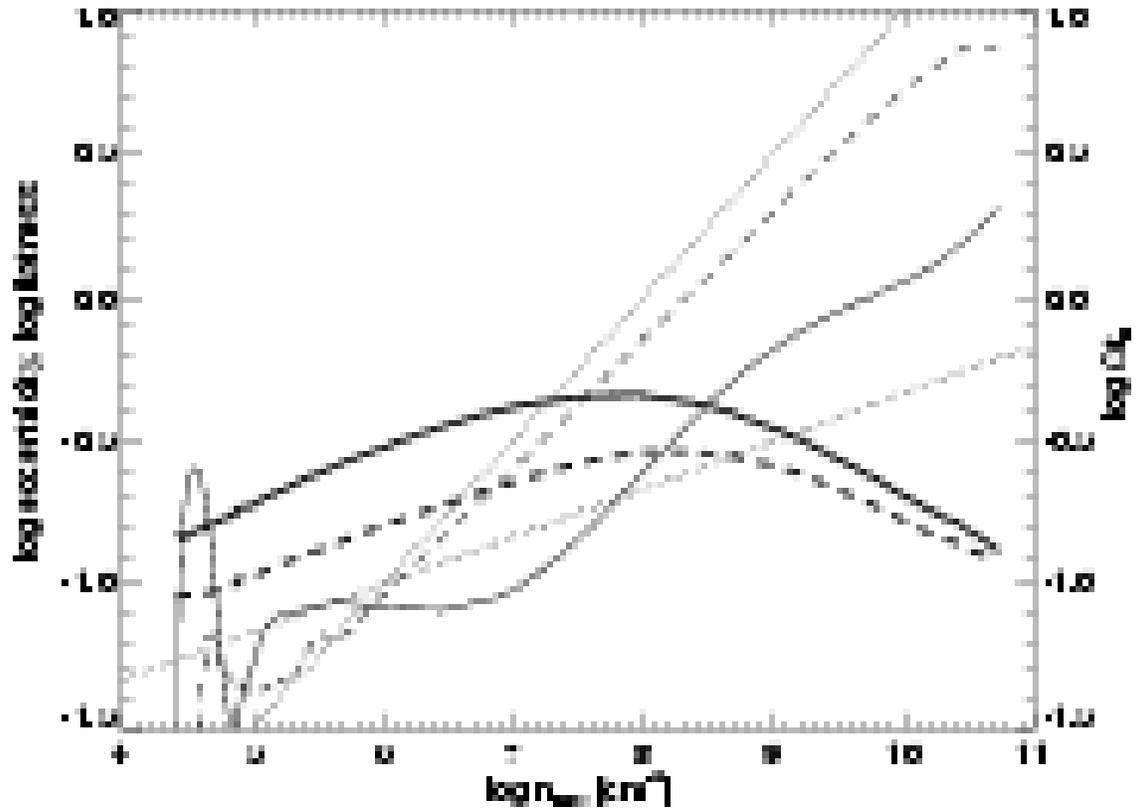}
\figcaption[barlength_plot0_b.eps]{
Same as Figure \ref{barlength_plot0.eps} but for a model of
{\it bar} type fragmentation followed by {\it satellite} type fragmentation,
($\Omega_0 t_{\rm ff}$, $\Omega_2 t_{\rm ff}$, $C$) = (0.2, 0.2, 0.15).
Dotted curves denote the relationships $\propto n_{\rm max}^{1/6}$ and
$n_{\rm max}^{1/2}$ for comparison.
\label{barlength_plot0_b.eps}
}
\end{figure}

Figure \ref{BE1.1c0.15_2.e-1_2.e-1_1.e-3.eps}{\it a} shows the cloud at the
end of the isothermal collapse phase, where it collapses to 
form a dense bar.  
Figure \ref{barlength_plot0_b.eps} shows the evolution of the eccentricity,
flatness and rotation of the central part of the cloud
in the isothermal collapse phase.  The central angular velocity in
unit freefall time, $\Omega_{0.5} t_{\rm ff}$ and $\Omega_{0.1} t_{\rm ff}$,
increases in proportion to $n_{\rm
max}^{1/6}$ and reaches
$\Omega_{0.5} t_{\rm ff} = 0.460$ at  $n_{\rm max} = 9.04 \times 10^7\,{\rm cm}^{-3}$ and
$\Omega_{0.1} t_{\rm ff} = 0.291$ at  $n_{\rm max} = 1.14 \times
10^8\,{\rm cm}^{-3}$.
Meanwhile, the cloud collapses almost spherically.
During $n_{\rm max} > 10^8\,{\rm cm}^{-3}$, $\Omega t_{\rm ff}$
decreases, and the flatness increases in proportion to 
$n_{\rm max}^{1/2}$ and
exceeds unity at $n_{\rm max} \simeq 10^8\,{\rm cm}^{-3}$.
The eccentricity also increases, although with significant oscillation. 
At the end of the isothermal collapse phase, $a_l$, $a_s$, and $a_z$
are 56.0~AU, 18.4~AU, and 3.84~AU, respectively.  
The long axis $a_l$ is 3.05 times longer than the short axis $a_s$.  

Figure \ref{BE1.1c0.15_2.e-1_2.e-1_1.e-3.eps}{\it b} shows the central cloud
at $t-t_{\rm cr}=210$~yr.
The bar-shaped, adiabatic core is surrounded by the isothermal disk.
At this stage, $a_l$, $a_s$, and $a_z$
are 30.4~AU, 2.80~AU, and 2.99~AU, respectively. The long axis $a_l$ is 10.9
times longer than the short axis $a_s$. 

Figure \ref{BE1.1c0.15_2.e-1_2.e-1_1.e-3.eps}{\it c} shows the fragmentation
of the narrow adiabatic bar at $t-t_{\rm cr}=510$~yr.
The bar is wound due to the differential rotation,
and develops two density peaks.   The separation between
these peaks is 12.0~AU. 
Figure \ref{BE1.1c0.15_2.e-1_2.e-1_1.e-3.eps}{\it d} shows the cloud at
$t-t_{\rm cr}=1.01 \times 10^3$~yr.
These density peaks merge to form a central core surrounded by
an adiabatic disk with spiral arms.
The spiral arms are the remnants of the wound bar. 
The disk is supported by centrifugal force and its radius is
approximately 20~AU at this stage. 
The disk radius increases due to the accretion of gas from the
infalling envelope, and by $t-t_{\rm cr} \simeq 1.5 \times 10^{3}$~yr, 
the disk has a radius of $\sim 40$~AU and deforms into a ring.
The ring is connected to the central kernel via the spiral arms, and 
the three satellite fragments form 
at the intersections of the ring and the spiral arms, as shown in 
Figure \ref{BE1.1c0.15_2.e-1_2.e-1_1.e-3.eps}{\it e}. 

Figure \ref{BE1.1c0.15_2.e-1_2.e-1_1.e-3.eps}{\it f} shows the 
stage of $t-t_{\rm cr}=1.25 \times 10^3$~yr (the last stage), in which five
fragments can be seen.  One of the two central fragments is formed
by subsequent {\it satellite} type fragmentation.  These fragments
rotate in a close orbit of 17.1~AU.
The masses of the fragments of the tight binary at the center are
$M_{13}= 1.6  \times 10^{-2} \, M_\odot$ and 
$1.7 \times 10^{-2}\, M_\odot$. 
The masses of the other fragments are
$M_{13}= 9.4 \times 10^{-3}\,M_\odot$,
$5.8 \times 10^{-3}\,M_\odot$, and 
$1.8 \times 10^{-3}\,M_\odot$ 
from inner to outer.

\subsubsection{Dumbbell type fragmentation}
\label{sec:dumbbell}

{\it Dumbbell} type fragmentation takes place in models with
large $\Omega_2 t_{\rm ff}$ and large $C$.
The cloud collapses to form a
dumbbell-shaped dense cloud having two density peaks
at the end of the isothermal collapse phase.  The dumbbell shape is
a hybrid of the ring and bar forms.
Although each of the density peaks evolves into a fragment, 
the fragments often merge as in {\it bar} type fragmentation.
{\it Satellite} fragments form at a later stage whenever {\it dumbbell} type
fragmentation occurs. 

\begin{figure}
\epsscale{0.9}
\plotone{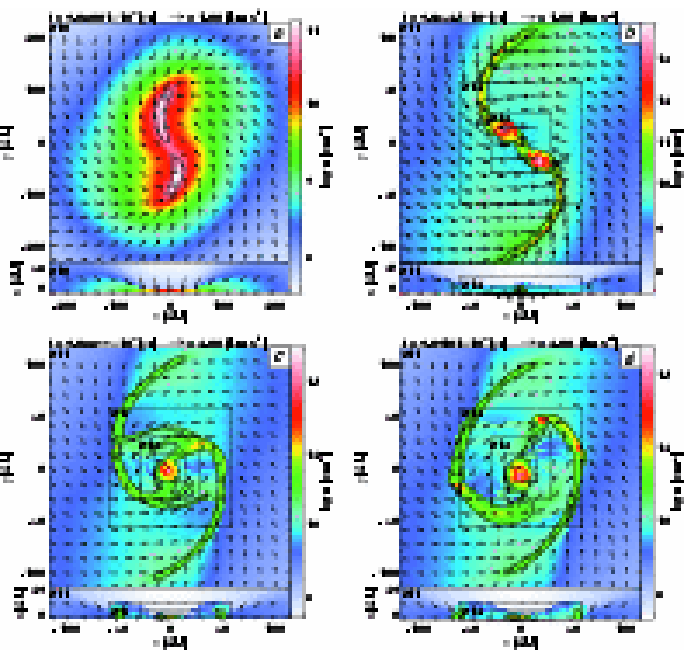}
\figcaption[BE1.1c0.5_2.e-1_2.e-1_1.e-3.eps]{
Same as Figure \ref{BE1.1c0_3.e-2_3.e-2_1.e-3.eps} but for a model of
{\it dumbbell} type fragmentation followed by
{\it satellite} type fragmentation,
$(\Omega_0 t_{\rm ff},\,\Omega_2 t_{\rm ff}, \, C)
=(0.2,\,0.2,\, 0.5)$.
\label{BE1.1c0.5_2.e-1_2.e-1_1.e-3.eps}
}
\end{figure}

\begin{figure}
\epsscale{0.9}
\plotone{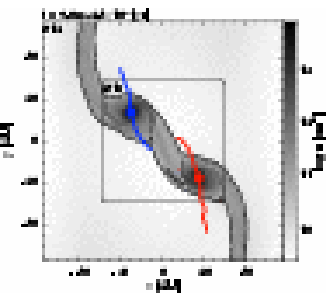}
\figcaption[BE1.1c0.5_2.e-1_2.e-2_1.e-3_orbit.eps]{
Same as Figure \ref{BE1.1c0_5.e-2_0.e+0_1.e-3_orbit.eps} but for a model of
{\it dumbbell} type fragmentation followed by
{\it satellite} type fragmentation,
$(\Omega_0 t_{\rm ff},\,\Omega_2 t_{\rm ff}, \, C)
=(0.2,\,0.2,\, 0.5)$.
\label{BE1.1c0.5_2.e-1_2.e-1_1.e-3_orbit.eps}
}
\end{figure}

Figure \ref{BE1.1c0.5_2.e-1_2.e-1_1.e-3.eps}
show the model of  $(\Omega_0 t_{\rm ff},\,\Omega_2 t_{\rm ff},\, C)
=(0.2,\,0.2,\, 0.5)$, as a typical model of {\it dumbbell} type fragmentation followed by
{\it satellite} type fragmentation.
The evolution of {\it dumbbell} type fragmentation is similar to that
of {\it bar} type fragmentation.
Figure \ref{BE1.1c0.5_2.e-1_2.e-1_1.e-3.eps}{\it a} shows the dense
dumbbell-shaped cloud at the beginning of the accretion phase
($t-t_{\rm cr}=151 $~yr).
The two density peaks evolve into self-gravitationally bounded fragments, as
shown in  Figure \ref{BE1.1c0.5_2.e-1_2.e-1_1.e-3.eps}{\it b} 
($t-t_{\rm cr} = 673$~yr),
surrounded by the isothermal disk.  
The fragments then merge to form a central kernel.
Figure \ref{BE1.1c0.5_2.e-1_2.e-1_1.e-3_orbit.eps} shows the loci of the
fragments during the merger. 

After the merger, the evolution of {\it dumbbell} type fragmentation
is very similar to that of {\it bar} type.
The central kernel is surrounded by a rotation-supported disk, which grows in
radius by the accretion of gas from the infalling envelope. 
Figure \ref{BE1.1c0.5_2.e-1_2.e-1_1.e-3.eps}{\it c} shows the disk
at $t-t_{\rm cr}=1.03 \times 10 ^3 $~yr.  The disk radius increases 
to $\simeq 40$~AU at this stage and deforms into a ring.
Figure \ref{BE1.1c0.5_2.e-1_2.e-1_1.e-3.eps}{\it d} shows the last
stage.
Satellite fragments form at the intersections of the
ring and the spiral arms as in model shown in Figure
\ref{BE1.1c0.15_2.e-1_2.e-1_1.e-3.eps}{\it f}.

\subsection{Classification of Fragmentation Processes}

\begin{figure}
\epsscale{0.9}
\plotone{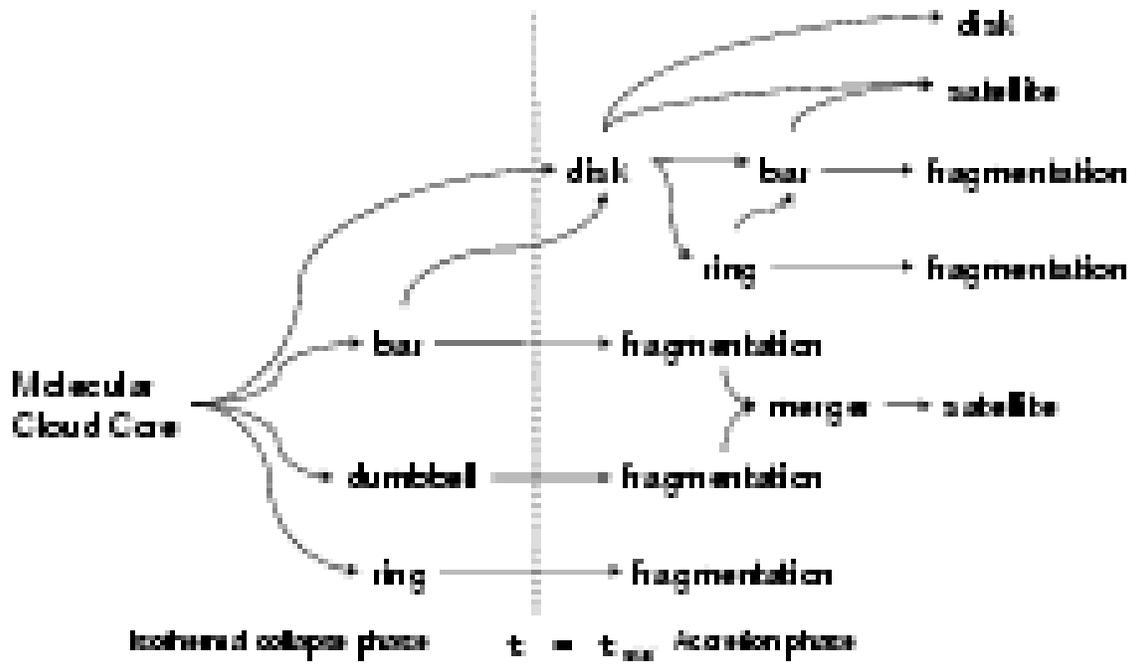}
\figcaption[classification.eps]{
Branching of models.
\label{classification.eps}
}
\end{figure}

The seven types of collapse
and fragmentation described above: {\it disk}, {\it disk-bar}, {\it ring-bar}, {\it
satellite}, {\it ring}, {\it bar}, and {\it dumbbell} types, are
summarized as a means of classification.
Figure \ref{classification.eps} shows the branching of these
fragmentation types schematically. In the isothermal collapse phase, cloud collapse is
classified into four main types: {\it disk}, {\it bar}, {\it
dumbbell}, and {\it ring} types.  {\it Disk} type collapse is subdivided
into {\it disk}, {\it satellite}, {\it bar}, and {\it ring} in the
accretion phase.


For all types, fragments (self-gravitationally confined clumps) form
only in the accretion phase.  
The isothermal collapse phase is so short that the cloud deforms into
a disk, bar, dumbbell, or ring-shape, but does not fragment.

\begin{figure}
\epsscale{0.9}
\plotone{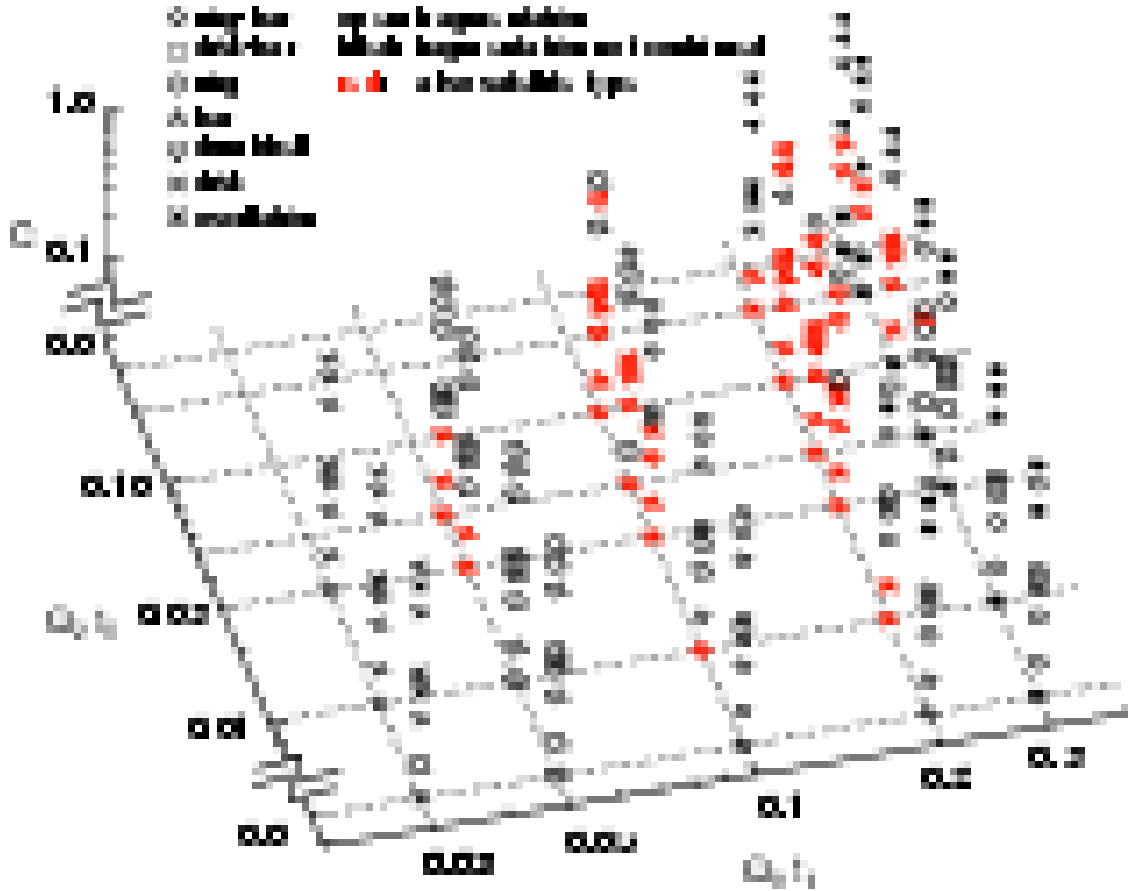}
\figcaption[summary3d.eps]{
Types of collapse and fragmentation in the three-dimensional phase
space of ($\Omega_0 t_{\rm ff}$, $\Omega_2 
t_{\rm ff}$, $C$). 
Red symbols denote the models of {\it satellite} type
fragmentation. For the models indicated by filled symbols, 
deformation of the central cloud could be followed,
but fragmentation could not, or survival of fragments cloud not be
confirmed in the simulation. 
\label{summary3d.eps}
}
\end{figure}

Figure \ref{summary3d.eps} summarizes domain of each type of collapse
and fragmentation in three-dimensional phase space ($\Omega_0 t_{\rm ff}$, $\Omega_2
t_{\rm ff}$, $C$).  The models with $\Omega_0 t_{\rm ff} \leq 0.03$ 
exhibit {\it disk} type collapse ($\times$) except for one model,
while almost all models with $\Omega_0 t_{\rm ff} \geq 0.05$ undergo
fragmentation. Some exceptional models exhibit {\it oscillation} ($\ast$).
The other parameters, $\Omega_2 t_{\rm ff}$ and $C$, specify
the type of fragmentation. 

The red symbols denote the models exhibiting {\it satellite} type
fragmentation.  The red cross ($\times$) denotes the model
exhibiting {\it disk} type
collapse followed by {\it satellite} type fragmentation, as shown in
\S\ref{sec:disk}.  Similarly the red triangle ($\triangle$)
denotes the model of {\it bar} type fragmentation followed by 
{\it satellite} type fragmentation.
Almost all the models proceeds to {\it satellite} type fragmentation 
when the bar mode $\Omega_2 t_{\rm ff}$ of the the initial cloud is
significant. 

Fragmentation could not be confirmed for the models indicated by
filled symbols. Almost all of these models have either a long bar, long dumbbell,
or large ring in the beginning of the accretion phase.  For the models
indicated by filled triangles ($\blacktriangle$) and inverted
triangles ($\blacktriangledown$), 
both the long bar and dumbbell are likely to fragment but
could not be confirmed due to  violation of the Jeans condition before
fragmentation.  From comparison with {\it bar} and {\it dumbbell}
type fragmentation as shown in \S\ref{sec:bar} and \S\ref{sec:dumbbell}, 
these fragments appear to merge, and
{\it satellite} type fragmentation should follow the merger.
Similarly, for models indicated by filled circles ($\bullet$), 
the formation of a ring but could be followed but subsequent
fragmentation cloud not.

\section{Discussion}
\label{sec:discussion}
\subsection{Collapse, Fragmentation, Survival, and Merger}

In this subsection, the fate of the collapsing clouds is discussed in detail.

A cloud fragments whenever the cloud collapses and 
the initial rotation is faster
than $\Omega_0 t_{\rm ff} \simeq 0.05$, independent of the other
parameters $\Omega_2$ and $C$.  The cloud collapses to
form a flat disk in the isothermal collapse phase, and
fragmentation of the flat disk is occurs by 
{\it disk-bar}, {\it ring-bar}, {\it ring}, {\it bar}, {\it
dumbbell}, and {\it satellite} types.  
When $\Omega_0 t_{\rm ff} \lesssim 0.05$, 
a flat disk forms in the accretion phase.
The flat disk formed in the isothermal collapse phase fragments, whereas
that formed in the accretion phase does not.

The criterion for fragmentation is given by the initial angular velocity.
The critical angular velocity, $\Omega_0 t_{\rm ff} \simeq 0.05$,
is evaluated in terms of spin-up of the collapsing cloud.
An isothermal cloud in runaway collapse spins up in
proportion to $\Omega_c t_{\rm ff} \propto \rho_c^{1/6}$, where
$\Omega_c$ denotes the angular velocity at the center
\citep{HanawaNakayama97}.  After the cloud changes its shape from a
sphere to a disk, the angular velocity becomes saturated at $\Omega_c
t_{\rm ff} \simeq 0.5$ \citep*{Matsumoto97,MH99}.  These quantities represent
a good index of disk formation. 

As shown in \S \ref{results}, the angular velocities $\Omega_{0.5}
t_{\rm ff}$ in models of {\it disk}, {\it disk-bar}, {\it ring-bar},
and {\it bar} types have maximum values of 0.0973, 0.452, 0.381, 0.460,
respectively, in the isothermal collapse phase.  When $\Omega_{0.5}
t_{\rm ff}$ is close to 0.5 ({\it disk-bar}, {\it ring-bar}, and {\it
  bar} types), the model forms a disk in the isothermal
collapse phase and fragments in the later stages.  
The small difference
between $\Omega_c$ and $\Omega_{0.5}$ is due to differential
rotation in the core.  The angular velocity is considerably smaller in
the rest of the models ({\it disk} type). 

Applying these quantities, the condition for formation of
a disk in the isothermal collapse phase can be evaluated as
\begin{equation}
\Omega_0 t_{\rm ff} \gtrsim 0.5 \left(\frac{\rho_c}{\rho_{\rm cr}}\right)^{1/6}
= 0.045\;.
\label{eq:criterion}
\end{equation}
This condition is consistent with the our simulations.
%

The formation of a flat disk in the isothermal
phase depends on the rotation of the central cloud, but not on the rotation
law specified by the parameter $C$.  In the isothermal
collapse phase, the central velocity and density become more important
to cloud collapse as the cloud shrinks \citep{Matsumoto97}.
During the isothermal (runaway) collapse phase, the mass of 
the central cloud decreases if defined as the mass
contained in the isodensity sphere of $ \rho _{\rm max} / 2 $.
The mass of the central cloud is only 0.01 $M _\odot$ at the end of the isothermal
phase.  Since fragmentation takes place in the central 0.01 $M _\odot $, 
the density and velocity thereof are important.   The
density and velocity in the envelope have little effect on the
initial fragmentation, but is involved in evolution of the fragments in
the accretion phase.

In the literature, fragmentation of clouds is typically discussed in
terms of the parameters $\alpha$ and $\beta$ \citep[e.g.,][]{Bodenheimer01},
which are evaluated by volume integration of energy for the entire 
cloud.  As described in equation (\ref{eq:beta}), $\beta$ depends not
only on the central rotation $\Omega_0$, but also on the rotation in
the envelope.  As shown in our models, the epoch of flat disk
formation depends solely on $ \Omega _0 $ and is affected little by
rotation in the envelope.  Thus, the central angular velocity
$\Omega_0$, rather than $\beta$, describes the formation of the flat
disk.

The cloud fragments when it satisfies the criterion of equation (\ref{eq:criterion}).  
The number of fragments and their orbital angular
momentum then determines whether the fragments merge or survive.  When the cloud
fragments into three or more fragments, 
many of the fragments survive, as shown in {\it ring-bar}, {\it satellite}
and {\it ring} types. 
On the other hand, when the cloud fragments into only two fragments,
the fate of the fragments depends on their orbital angular momenta: 
fragments with high orbital angular momentum survive (e.g., {\it
  disk-bar} type), while fragments with low orbital angular
momentum merge (e.g., {\it bar} and {\it dumbbell} types).  In the latter
case, satellite fragments form after the merger.


The orbital angular momentum of the fragment depends on the timing of
deformation into the bar-shape.  When the cloud deforms into a bar in the
isothermal collapse phase, the bar does not have sufficient spin
angular momentum to be supported by the rotation.  In the isothermal
collapse phase, the cloud undergoes runaway collapse and is never
supported by the rotation \citep{saigo98}.  Thus, the orbital motions
of the fragments are also never supported by the rotation.  Furthermore,
the bar loses angular momentum via gravitational torque in the
accretion phase.
In the model shown
in \S\ref{sec:dumbbell} ({\it dumbbell} type), each fragment has a
specific orbital angular momentum of $  6.5 \times 10^{18}\, {\rm
  cm}^{2}\,{\rm s}^{-1}$ at the stage shown in Figure
\ref{BE1.1c0.5_2.e-1_2.e-1_1.e-3.eps}{\it b}. 
This specific orbital angular momentum is only 20\% of that
required for support by rotation at $R=20$~AU.
On the other hand, when the cloud deforms into a bar in the accretion phase,
the bar is supported by the rotation (e.g., {\it disk-bar} type).
The adiabatic disk accretes gas with high specific angular
momentum and is already supported by rotation prior to
its fragmentation.

\subsection{Comparison with Earlier  Numerical Simulations}

{\it Bar} and {\it dumbbell} type
fragmentation were also seen in the numerical simulation of
\citet{Boss00}.  They followed the evolution of clouds having an initial
Gaussian density profile by three types of approximations;
isothermal equation of state,  barotropic equations of state, and the Eddington
approximation of radiative transfer. 
The barotropic model of \citet{Boss00} exhibits {\it bar} type fragmentation.
The density profiles shown in their Figs.~5{\it c} and 5{\it d}
are similar to those shown in 
Figures \ref{BE1.1c0.15_2.e-1_2.e-1_1.e-3.eps}{\it b}
and \ref{BE1.1c0.15_2.e-1_2.e-1_1.e-3.eps}{\it d} here. 
Using AMR code, they followed the formation of a bar, fragmentation
of the bar, and merger of the fragments. 
This evolution resembles that of the model shown in \S\ref{sec:bar},
although the initial condition is quite different. 
They terminated the
calculation at the stage of adiabatic disk formation after the merger,
and the stage of {\it satellite} type fragmentation was not shown. 
{\it Dumbbell} type fragmentation is also seen in their Fig.~2{\it a},
which is quite similar to Figure
\ref{BE1.1c0.5_2.e-1_2.e-1_1.e-3.eps}{\it a} here.  {\it Dumbbell} type
fragmentation in their study was computed under the Eddington approximation.

The initial cloud of \citet{Boss00} is defined by $\alpha = 0.26$ whereas
$\alpha = 0.765 $ in the present study.  
Their cloud was thus colder, or in other words, more massive. 
Despite this difference, the fragmentation is very similar.  
The same mechanism of deformation and fragmentation therefore appears
to be valid over a wide range of $\alpha$.

{\it Satellite} type fragmentation has also been seen in many other
simulations \citep{Bonnell94,Burkert97,Bate02}.  \citet{Bonnell94}
followed the fragmentation in the second collapse, in which the first
core collapses to form the second core.  In their simulations,
satellite fragments form through interaction of the spiral arms.
The satellite fragments form in the same manner both in their
and our simulations,
even though different situations are considered.
\citet{Burkert97} followed the collapse and
fragmentation of molecular cloud cores similar to this paper.
They also computed the formation of satellite fragments and followed
their orbits using a nested grid.  
It was not explicitly mentioned whether the nested grid
simulation satisfies the Jeans condition.
\citet{Burkert97} and \citet{Bate02} confirmed the result by
independent simulations using an SPH code.

\subsection{Application to Formation of Binary and Multiple Stars}

Some observations have indicated the rotation of molecular cloud cores.
\citet{Goodman93} found that 29 of 43 molecular clouds had significant
velocity gradient, corresponding to rigid rotation of $2\times10^{-3}
< \beta < 1.4$ with typical values of $\beta \sim 0.02$.  
These
quantities correspond to $0.047 < \Omega_0 t_{\rm ff} < 1.25$ with
typical values of $\Omega_0 t_{\rm ff} \sim 0.15$ for models of
$(\Omega_2 t_{\rm ff}, \, C) = (0,\,0)$.
Unfortunately, the observations were not of sufficient accuracy to
specify the rotation law. 
Our simulations show that collapsing clouds having an initial
rotation of  $\Omega_0 t_{\rm   ff} \gtrsim 0.05$  fragment, which is 
consistent with observed high binary frequency.

Molecular cloud cores have internal motion often interpreted as
turbulence.  The internal motion should reflect the superposition of
various modes of velocity perturbations.  When the bar mode of a cloud
is a significant, the cloud undergoes {\it satellite} type
fragmentation as shown in our simulations.  Therefore, {\it satellite} type
fragmentation may be dominant.
The satellite fragments merge and scatter while
accreting gas.
Consequently, the satellites will have various binary separations. 
This may explain the wide range of separation for 
young and main sequence binaries \citep[e.g.,][]{Mathieu94}.

It has recently been suggested that brown dwarfs may be formed by ejection of
the seeds of stars from a parent cloud core \citep{Reipurth01,Bate02}.  
{\it Satellite}
type fragmentation might be a corresponding case.
In many case of {\it satellite} type fragmentation, 
three or more fragments are formed.  In these multiple systems, it is possible that
a close encounter will eject the fragment from the cloud center. The satellite
fragment has speed of $\sim 1 \,{\rm km}\,{\rm s}^{-1}$ at the last
stage of the model of 
($\Omega_0 t_{\rm ff}$, $\Omega_2 t_{\rm ff}$, $C$) = (0.2, 0.2, 0.15)
(for a fragment shown in the right side of Figure
\ref{BE1.1c0.15_2.e-1_2.e-1_1.e-3.eps}{\it f}). 
The velocity would be reduced substantially before
ejection by the gravity of the molecular cloud core.
The gravitational potential is evaluated to be
$ \psi \approx 2 c _{\rm s} ^2 \ln r $, and
the ejection speed would be $v_{\rm escape} \sim 1 \,{\rm km}\,{\rm
s}^{-1}$. 
The ejected satellite fragment would have a velocity
of the order of the escape speed if it exits.

\section{Summary}
\label{sec:summary}
The collapse and fragmentation of molecular cloud cores was investigated
for the case that the initial cloud is almost in equilibrium, 
focusing on the effects of rotation speed, rotation law, and bar mode
perturbation.  The main results are summarized as follows.

A cloud 1.1 times denser than the critical Bonnor-Ebert sphere
fragments when rotation of the initial cloud is slowly enough to allow
collapse, but still significant, i.e., $\Omega_c t_{\rm ff} \gtrsim 0.05 $.  
The latter condition gives rise to the formation of a flat
disk in the isothermal collapse phase. This condition is independent
of both the initial amplitude of the bar mode and the initial rotation law.

Six types of fragmentation were identified: {\it disk-bar}, {\it ring-bar}, {\it
satellite}, {\it bar}, {\it ring}, and {\it dumbbell} types. The 
type of fragmentation depends on the initial amplitude of the bar mode
and the initial rotation law.  The fragments formed via {\it bar} or {\it
dumbbell} types fragmentation merge due to their low angular momenta,
and new fragments form via {\it satellite} type fragmentation.  In other
words, a cloud forms satellite fragments whenever the bar mode of the initial cloud
is appreciable amplitude. Merger and close encounter of the satellite
fragments may result in the wide range of the binary separation.

\acknowledgments

The authors thank S.~Inutsuka, T.~Tsuribe, and K.~Saigo for valuable discussion.
Numerical computations were carried out on VPP5000 at
the Astronomical Data Analysis Center of the National Astronomical
Observatory, Japan, which is an inter-university research institute of
astronomy operated by 
the Ministry of Education, Culture, Sports, Science and
Technology, Japan (MEXT).
This study was financially supported
in part by Grants-in-Aid for the Encouragement of Young Scientists (12740123, 14740134), 
for Scientific Research on Priority Areas (A) (13011204) from MEXT, 
and for Scientific Research (C) (13640237) from the Japan Society of Promotion of Science (JSPS),

\clearpage


\clearpage

\clearpage

\clearpage

\clearpage


\begin{deluxetable}{lll}
\tabletypesize{\scriptsize}
\tablecaption{High-resolution models \label{table:model}}
\tablewidth{0pt}
\tablehead{
\colhead{$\Omega_0 t_{\rm ff}$} & \colhead{$\Omega_2 t_{\rm ff}$}   & \colhead{$C$} 
}
\startdata
0.03 & 0.03 & 0.0  \\
0.05 & 0.0  & 0.0  \\
0.1  & 0.0  & 0.0  \\
0.1  & 0.03 & 0.0  \\
0.1  & 0.1  & 0.0  \\
0.1  & 0.05 & 0.0  \\
0.2  & 0.2  & 0.15 \\
0.2  & 0.2  & 0.5  \\
0.2  & 0.2  & 1.0  \\
 \enddata
\end{deluxetable}

\end{document}